\documentclass[twocolumn,superscriptaddress,amsmath,amssymb,floats]{revtex4-1}
\usepackage[latin9]{inputenc}
\usepackage{color}
\usepackage{mathtools}
\usepackage{amsmath}
\usepackage{amssymb}
\usepackage{graphicx}
\usepackage{dsfont}
\usepackage{verbatim}
\usepackage{epstopdf}
\usepackage{verbatim}
\usepackage{float}
\usepackage[bookmarks=false,colorlinks=true,urlcolor=blue,citecolor=blue,linkcolor=blue]{hyperref}

\makeatletter

\@ifundefined{textcolor}{}
{%
 \definecolor{BLACK}{gray}{0}
 \definecolor{WHITE}{gray}{1}
 \definecolor{RED}{rgb}{1,0,0}
 \definecolor{GREEN}{rgb}{0,1,0}
 \definecolor{BLUE}{rgb}{0,0,1}
 \definecolor{CYAN}{cmyk}{1,0,0,0}
 \definecolor{MAGENTA}{cmyk}{0,1,0,0}
 \definecolor{YELLOW}{cmyk}{0,0,1,0}
}

\usepackage{epstopdf}
\usepackage{array}
\usepackage{mathrsfs}
\usepackage{dcolumn}
\usepackage{bm}

\newcolumntype{L}[1]{>{\raggedright\let\newline\\\arraybackslash\hspace{0pt}}m{#1}}
\newcolumntype{C}[1]{>{\centering\let\newline\\\arraybackslash\hspace{0pt}}m{#1}}
\newcolumntype{R}[1]{>{\raggedleft\let\newline\\\arraybackslash\hspace{0pt}}m{#1}}

\newcommand{\mbf}[1]{\mathbf{#1}}

\usepackage{tikz}
\usetikzlibrary{calc,intersections}
\usetikzlibrary{shadings}
\usepgflibrary{shadings}

\makeatother

\begin{document}

\title{Spiral magnetic order and topological superconductivity in a chain of magnetic adatoms on a two-dimensional superconductor}
\author{Morten H. Christensen}
\affiliation{Niels Bohr Institute, University of Copenhagen, DK-2100, Denmark}
\author{Michael Schecter}
\affiliation{Center for Quantum Devices, Niels Bohr Institute, University of Copenhagen, DK-2100, Denmark}
\author{Karsten Flensberg}
\affiliation{Center for Quantum Devices, Niels Bohr Institute, University of Copenhagen, DK-2100, Denmark}
\author{Brian M. Andersen}
\affiliation{Niels Bohr Institute, University of Copenhagen, DK-2100, Denmark}
\author{Jens Paaske}
\affiliation{Center for Quantum Devices, Niels Bohr Institute, University of Copenhagen, DK-2100, Denmark}

\begin{abstract}
We study the magnetic and electronic phases of a 1D magnetic adatom chain on a 2D superconductor. In particular, we confirm the existence of a `self-organized' 1D topologically non-trivial superconducting phase within the set of subgap Yu-Shiba-Rusinov (YSR) states formed along the magnetic chain. This phase is stabilized by incommensurate spiral correlations within the magnetic chain that arise from the competition between short-range ferromagnetic and long-range antiferromagnetic electron-induced exchange interactions, similar to a recent study for a 3D superconductor [M. Schecter \textit{et al.} Phys. Rev. B \textbf{93}, 140503(R) 2016]. The exchange interaction along diagonal directions are also considered and found to display behavior similar to a 1D substrate when close to half filling. We show that the topological phase diagram is robust against local superconducting order parameter suppression and weak substrate spin-orbit coupling. Lastly, we study the effect of a direct ferromagnetic exchange coupling between the adatoms, and find the region of spiral order in the phase diagram to be significantly enlarged in a wide range of the direct exchange coupling.
\end{abstract}
\maketitle

\section{Introduction}

The study of magnetic order in adatomic chains deposited on superconducting substrates has recently attracted widespread attention due to the ability of these systems to host Majorana bound states \cite{nadj-perge13,nadj-perge14,yazdani3,berlin,basel,pientka1,pientka2,pientka3,klinovaja13,braunecker13,vazifeh13,sarma1,heimes1,brydon-sau,reis14,ojanen1,hu15,braunecker15,schecter16}. The local moments of the adatoms induce Yu-Shiba-Rusinov (YSR) bound states within the superconducting gap \cite{yu,shiba,rusinov,balatsky,yazdani4,yazdani5}, thus constituting an effective Kitaev chain \cite{kitaev01} with long-range hopping and pairing amplitudes \cite{pientka1}. A topologically non-trivial phase is possible with the addition of a further crucial ingredient, namely an effective spatial variation of the local exchange field experienced by the electrons along the chain \cite{pientka1}. This can be achieved either by spin-orbit coupling (SOC) within the superconductor \cite{nadj-perge14,yazdani3,heimes1,brydon-sau}, or without SOC if the moments order into a magnetic spiral \cite{nadj-perge13,klinovaja13,braunecker13,vazifeh13,reis14,ojanen1,hu15,braunecker15,schecter16} (see also \cite{choy, martin,kjaergaard,braunecker10a}). In the latter case, spiral order is driven by electron-mediated indirect exchange interactions that in turn support topological superconductivity and give rise to the notion of self-organization.

The development of magnetic order in an adatom chain due to electron-induced exchange interactions has been studied analytically for both one- and three-dimensional superconductors \cite{klinovaja13, braunecker13, vazifeh13,schecter15,hu15,braunecker15,schecter16}. In one-dimensional (1D) conductors, adatom spiral order has been shown to arise from the RKKY interaction \cite{ruderman54,kasuya56,yosida57} due to the singular behavior of the susceptibility at $2k_F$ \cite{klinovaja13, braunecker13, vazifeh13,braunecker15}. Effects beyond the RKKY approximation were recently considered and also support the formation of spiral order away from points of commensurability, and for weak adatom-electron exchange coupling~\cite{schecter15,hu15}. The three-dimensional (3D) case was studied in Ref.~\onlinecite{schecter16} where it was found that spiral order indeed forms due to indirect exchange interactions, however, the mechanism is distinct from the 1D case since there is no $2k_F$ peak in the adatom susceptibility \cite{sarma1}. In 3D spiral order arises from the interplay between the shorter-ranged RKKY exchange, and the longer-ranged antiferromagnetic exchange due to singlet superconductivity \cite{schecter16}. 

In two-dimensional (2D) systems the existence of self-organized topological phases was established numerically for finite systems~\cite{reis14}, but the mechanism and conditions under which spiral magnetic order forms are not yet fully understood.
In addition, single YSR states were recently imaged in the layered superconductor 2H-NbSe$_2$~\cite{menard16}, which demonstrates how the effectively reduced dimensionality enhances the spatial extent of the YSR states. This is expected to lead to a larger YSR pairing hybridization, and thus to a relatively larger gap protecting the topological superconducting phase. 

In this paper we bridge the gap between the previous 2D numerical and 3D analytical calculations by providing comprehensive studies of the magnetic adatom and electronic ground states in a two-dimensional tight-binding model. We map out the magnetic phase diagram as a function of exchange coupling and electron chemical potential by minimizing the electron free energy within a classical spiral ansatz for the adatom chain. We find that the indirect exchange interactions generally follow behavior similar to 3D studies, favoring collinear order of the adatom chain in the normal state, while destabilizing ferromagnetism to spiral formation in the presence of superconductivity. This gives rise to a broad region of the phase diagram where the set of subgap YSR states exists in a topologically nontrivial superconducting phase with Majorana bound states. The exchange interaction along the diagonal (11) direction is distinct near half filling due to Fermi surface nesting. As a result, the effective dimensionality of the substrate is reduced, and the magnetic order along the chain exhibits $2k_{F}$ spiral order known from 1D systems. Furthermore, we ascertain the effects of a direct exchange interaction between adatoms, finding that even a substantial direct exchange term can promote spiral order in the chain. This is contrary to 3D systems, where a spiral state in general only occurs when the direct exchange interaction is smaller than the indirect exchange.

Lastly, we elucidate the differences between performing the calculations selfconsistently and non-selfconsistently for the local pairing potential. The two cases are found to be qualitatively the same, i.e.,  suppression of the local pairing potential near the adatom chain leads only to minor modifications of the magnetic order and subgap states. This modification is interpreted in terms of a lowering of the effective chemical potential for the subgap YSR states induced by the suppression of the local pairing potential.

The paper is organized as follows: In Section \ref{sec:model} we introduce the model and methods. In Section \ref{sec:exchange} we study the indirect exchange interactions between two adatom spins mediated by the electron gas, and determine the dependence on chemical potential and exchange coupling both along (10) and (11) directions. We proceed to consider chains of magnetic adatoms in Sec. \ref{sec:chain}. We present magnetic and topological phase diagrams for different values of the superconducting order parameter in the plane of exchange coupling and chemical potential. These indicate that spiral order on a chain along (10) is formed by a mechanism similar to the 3D case. In this section we also contrast the behavior of chains along (10) and (11), and reveal substantial differences that arise due to the anisotropic Fermi surface. Additionally, we discuss the effects of a ferromagnetic direct exchange between the adatoms. In Sec.~\ref{sec:selfconsistency} we perform a detailed comparison between selfconsistent and non-selfconsistent approaches, and find that the two approaches yield qualitatively similar results. We discuss the influence of substrate spin-orbit coupling in the Appendix. Conclusions and outlook are presented in Sec. \ref{sec:conclusions}.

\section{Model}\label{sec:model}

\begin{figure}[!t]
\centering
\includegraphics[width=\columnwidth]{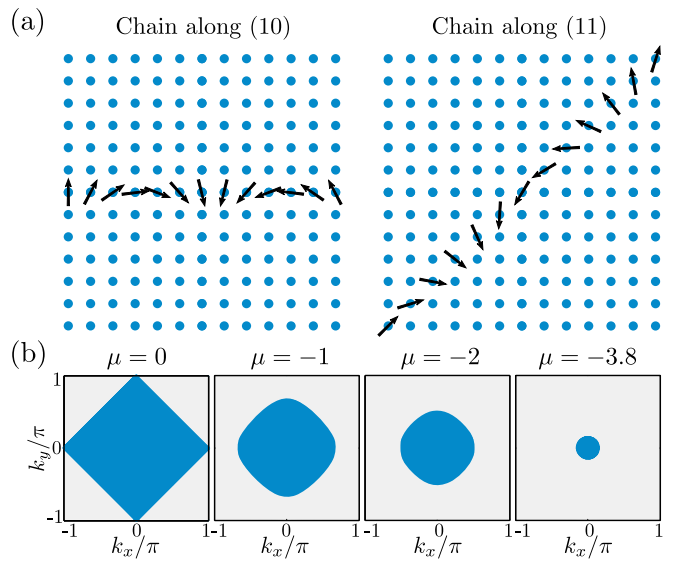}
\caption{\label{fig:system_config} (a) Schematic illustration of the system under consideration. In this paper we consider both the (10)-direction and the (11)-direction, for respectively 120$\times$71, and 81$\times$81 systems. (b) Fermi surface of the model in Eq.~(\ref{eq:h0}) for various values of the chemical potential.}
\end{figure}

To model the 2D superconducting substrate we use a tight-binding model with an on-site attractive interaction $V$ to stabilize superconductivity. The magnetic adatom potentials are assumed to be local and are arranged into a chain along either the (10) or (11) directions, depicted in Fig.~\ref{fig:system_config}(a). The Hamiltonian is
\begin{eqnarray}
	\mathcal{H} &=& \mathcal{H}_{0} + \mathcal{H}_{\text{SC}} + \mathcal{H}_{\text{imp}}\,,\label{eq:fullh}\\
	\mathcal{H}_{0} &=& -t\sum_{\substack{\langle ij \rangle \\ \alpha}} c^{\dagger}_{i\alpha}c_{j\alpha} - \mu\sum_{i\alpha}c^{\dagger}_{i\alpha}c_{i\alpha} \,,\label{eq:h0}\\
	\mathcal{H}_{\text{SC}} &=& - V\sum_{i}c^{\dagger}_{i\uparrow}c^{\dagger}_{i\downarrow}c^{\phantom{\dagger}}_{i\downarrow}c^{\phantom{\dagger}}_{i\uparrow}\,,\label{eq:hsc}\\
	\mathcal{H}_{\text{imp}} &=& J_{\text{imp}}\sum_{\substack{i\in \mathcal{I} \\ \alpha \beta}}\mbf{S}_{i}\cdot c^{\dagger}_{i\alpha}\boldsymbol{\sigma}_{\alpha\beta}c_{i\beta}\,,\label{eq:himp}
\end{eqnarray}
where $c^\dagger_{i\alpha},c_{i\alpha}$ are fermionic creation/annihilation operators with spin $\alpha$ and coordinate $i$, $\mu$ is the chemical potential, $\mathcal{I}$ is the set of adatom locations, $\langle \rangle$ signifies that the summation is taken over nearest-neighbors, and $\boldsymbol{\sigma}$ is the vector of Pauli matrices. We choose $t=1$ as the unit of energy and the lattice constant $a=1$ as the unit of length. The adatom spin is denoted by $\mbf{S}=S\hat{\mbf{n}}$ where $\hat{\mbf{n}}$ is a unit vector in the direction of the spin and $S$ is the length. Throughout the paper we work in the large spin (classical) approximation, $S \rightarrow \infty$, $J_{\text{imp}} \rightarrow 0$ with the product $J\equiv J_{\text{imp}}S=\text{const}$.
A mean-field decoupling in the Cooper channel is performed on the superconducting term Eq.~(\ref{eq:hsc}) resulting in
\begin{eqnarray}
	\mathcal{H}^{\text{MF}}_{\text{SC}} &=& -\sum_{i} \left[ \Delta_{i}c^{\dagger}_{i\uparrow}c^{\dagger}_{i\downarrow} + \text{h.c.} - \frac{|\Delta_i|^2}{V} \right] \,,
\end{eqnarray}
where the superconducting order parameter is obtained \textit{via} the selfconsistency equation
\begin{eqnarray}
	\Delta_{i} = V \langle c_{i\downarrow}c_{i\uparrow} \rangle\,. \label{eq:self-consistency_eq}
\end{eqnarray}
The Fermi surfaces (with $V=J=0$) for various representative values of the chemical potential are shown in Fig.~\ref{fig:system_config}(b). The dispersion inherits the point group symmetries of the square lattice, and a circular Fermi surface with quadratic dispersion is only achieved near the bottom of the band. We note that the tight-binding model has a finite band-width $W=8t$ and is particle-hole symmetric around $\mu=0$, implying that our results do not depend on the sign of $\mu$.
For purposes of determining both indirect exchange interactions and the magnetic order of the adatoms, we consider the thermodynamic potential $\Omega$ at zero temperature
\begin{eqnarray}\label{eq:therm-dyn}
	\Omega &=& \langle \mathcal{H} \rangle\,.
\end{eqnarray}
Below we study $\Omega\{ \hat{\mbf{n}}_{i} \}$ for different adatom configurations $\hat{\mbf{n}}_{i}$ and determine the magnetic ground state for a chain of adatoms by minimizing $\Omega$. To obtain an iterative selfconsistent solution to the Hamiltonian~(\ref{eq:fullh}) we solve Eq.~(\ref{eq:self-consistency_eq}) for a given $V$ and $\Delta_i$ and iterate until the difference between consecutive solutions is $<10^{-3}$ at each site. This procedure includes the feedback of the adatoms on the superconducting order parameter and suppresses it in the proximity of the chain, as depicted in Fig.~\ref{fig:local_SC_OP_suppression} below. This leads to the well-known $\pi$--phase shift of the superconducting order parameter at the adatom site~\cite{salkola97,balatsky,flatte97}. As will be made clear in Sec.~\ref{sec:selfconsistency} this effect has no qualitative impact on the magnetic order along the chain, or the subgap YSR states. In Secs.~\ref{sec:exchange} and \ref{sec:chain} we therefore simplify the calculations and use the non-selfconsistent approximation.

\section{Weak exchange interactions}\label{sec:exchange}

\begin{figure}[t]
\centering
\includegraphics[width=0.82\columnwidth]{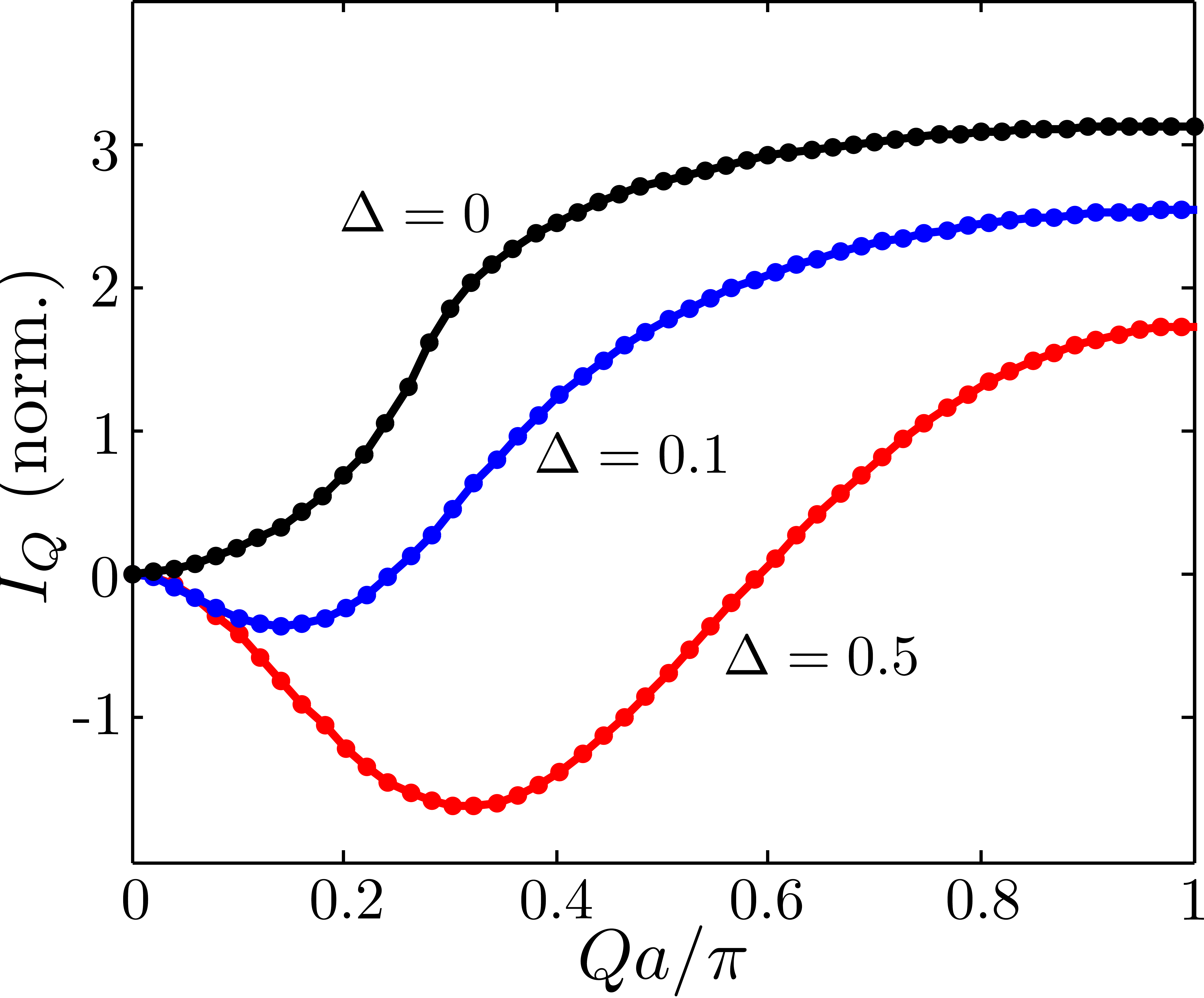}
\caption{\label{fig:I_of_q} (Color online) The function $I_Q$ in Eq.~(\ref{eq:Heis}) calculated from Eq.~(\ref{eq:fullh}) to leading order in $J$ at $\mu=-3.8$ and for different values of $\Delta$. The two adatoms are arranged along the (10) direction with lattice spacing $a_{\rm ad}=a$. In the normal state ($\Delta=0$), the magnetic ground state of the adatom chain is a ferromagnet ($q=0$). The presence of superconductivity leads to a spiral magnetic ground state $(q\neq0)$ of the adatom chain. Here we assume a homogeneous pairing potential $\Delta_i=\Delta$. The  effects of the selfconsistency condition, Eq.~(\ref{eq:self-consistency_eq}), are addressed in Sec.~\ref{sec:selfconsistency}.}
\end{figure}

To understand the magnetic phases of the adatom chain, we first consider the case where the adatom spacing $a_{\rm ad}$ is larger than the inverse Fermi wavevector, $k_F a_{\rm ad}\gg1$, and the exchange coupling to electrons is weak. The indirect exchange coupling between adatoms can then be computed perturbatively in $J$, resulting in an effective Heisenberg model for the adatoms given by

\begin{eqnarray}
\label{eq:Heis}
	\mathcal{H}_{\text{Heis}}=\sum_{i,j}I_{(i-j)}\mbf{S}_{i}\cdot\mbf{S}_{j}=\sum_Q I_Q |\mbf{S}_Q|^2,
\end{eqnarray}
where the second equality is written in the momentum representation for an infinite ring. For a chemical potential near the band bottom, the leading order in $J$ indirect exchange coupling between adatoms separated by distance $r$ is $(\hbar=1)$
\begin{align}
\label{eq:I}
I_r\propto J^2e^{-\frac{2r}{\xi}}\left[-\frac{v_F}{2\pi r^2}\mathrm{sin}(2k_Fr)+\frac{\Delta}{r}\,\mathrm{sin}^2(k_Fr+\pi/4)\right],
\end{align}
where $v_F$ is the Fermi velocity and $\xi=v_F/\Delta$ is the coherence length of the superconductor. The first term in the square brackets of Eq.~\eqref{eq:I} is the well-known Rudermann-Kittel-Kasuya-Yosida (RKKY) interaction~\cite{ruderman54,kasuya56,yosida57} mediated by a 2D electron gas \cite{sarma1,aristov1}. The second term is purely antiferromagnetic and arises from singlet superconducting correlations that disfavor the pair-breaking effect of a polarized exchange field \cite{aristov2,abrikosov,yao14}. The magnetic ground state $\mbf{S}_q$ can be determined to second order in $J$ by finding the minimum Fourier component of the exchange interaction $I_Q$, see Fig.~\ref{fig:I_of_q}. Here we label a generic magnetic wavevector by $Q$, and denote the configuration minimizing the thermodynamic potential by $q$.

\subsection{Exchange interactions along the (10) direction}\label{sec:exchange_10}

In the normal state $(\Delta=0)$ the magnetic ground state calculated from Eq.~(\ref{eq:I}) is a ferromagnet ($q=0$) in the range $n<k_Fa_{\rm ad}/\pi<n+1/2$ with integer $n$ and an antiferromagnet ($q=\pi/a$) otherwise. In the presence of superconductivity the antiferromagnet is stable, while the ferromagnet becomes unstable to the formation of a spiral with finite $q\neq0$. Indeed, for $\Delta\neq0$, $\xi^{-1}\ll Q\ll\pi/a_{\rm ad}$ the exchange interaction scales like $I_Q\propto \cot (k_{\rm F} a_{\rm ad}) v_FQ^2/k_F  -\Delta\mathrm{ln}(Qa_{\rm ad})/(k_Fa_{\rm ad})$, so that the ground state wavevector is shifted from zero to  $q\propto\sqrt{\Delta}$. 

This magnetic instability is akin to the Anderson-Suhl transition in 2D and 3D spin lattices \cite{abrikosov,aristov2,anderson-suhl} and results from two competing ordering mechanisms having different strengths and effective ranges: ferromagnetism from the RKKY exchange and antiferromagnetism due to superconductivity. The development of spiral order due to the presence of superconductivity is illustrated in Fig.~\ref{fig:I_of_q}. The spiral formation of a 1D spin chain on a 3D superconductor was recently demonstrated in Ref.~\onlinecite{schecter16}, where the wavevector scales as $q\propto\Delta$ in contrast to $q\propto\sqrt{\Delta}$ found above. One can easily generalize this result to a superconductor/adatom lattice of arbitrary dimensions to find $q\propto \Delta^{1/(3-D^*)}$, where $0\leq D^*\leq 2$ is the codimension of the adatom lattice in the $s$-wave superconductor (the case of nodal $d-$wave superconductors requires a separate analysis \cite{aristov2}). The famous Anderson-Suhl scaling $q\propto \Delta^{1/3}$ \cite{anderson-suhl} is obtained only when the adatom lattice and superconductor have the same dimension, $D^*=0$. This indicates that for the adatom chain, the influence of superconductivity on the magnetic order is substantially enhanced for a 2D substrate as compared to a 3D substrate.

\begin{figure}[t]
\centering
\includegraphics[width=\columnwidth]{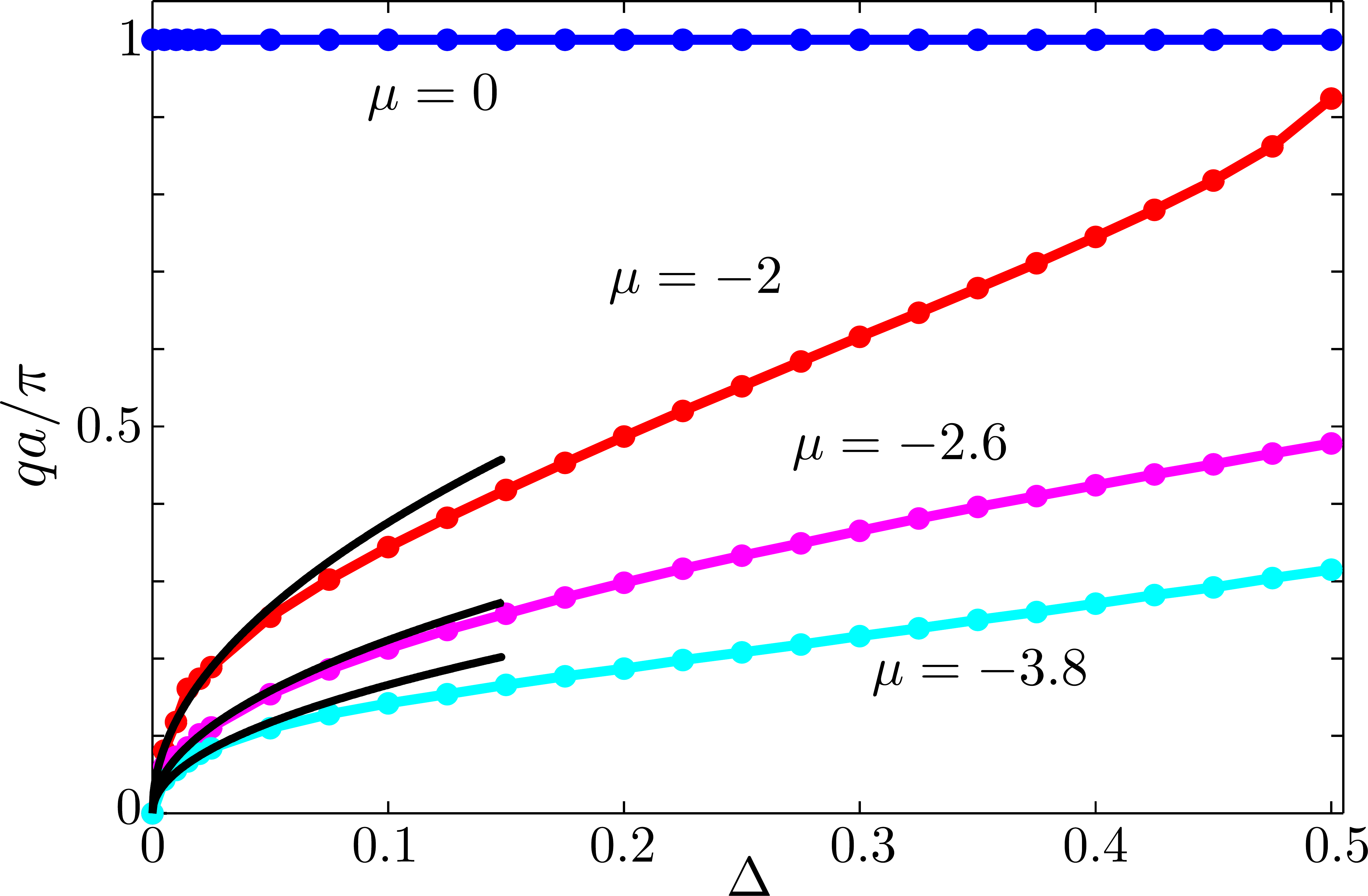}
\caption{\label{fig:q_as_Delta} (Color online) Ground state wavevector $q$ of an adatom chain along the (10) direction ($a_{\rm ad}=a$) calculated from Eq.~(\ref{eq:fullh}) to leading order in $J$ as a function of $\Delta$ for different values of $\mu$.  The data points for $qa/\pi
\lesssim0.2,\,\Delta\lesssim 0.05$ are well-fit by the form $q\propto\sqrt{\Delta}$ (black lines), as predicted from the analysis of Eq.~(\ref{eq:I}).}
\end{figure}

We illustrate the dependence of $q$ on $\Delta$ in Fig.~\ref{fig:q_as_Delta}, calculated for the model of Eq.~(\ref{eq:fullh}) to leading order in $J$, for a dense set of adatoms along the (10) direction ($a_{\rm ad}=a$). The black lines illustrate the proposed square-root behavior of $q(\Delta)$. The dependence of $q$ on $\mu$ can be traced back to Eq.~(\ref{eq:I}). For $\mu$ close to half filling we have $1/2< k_F a_{\rm ad} / \pi < 1$ and the resulting state is antiferromagnetic. For $\mu=-2$, we find $k_F a_{\rm ad} /\pi =1/2$ and there is a first order transition from an antiferromagnetic to a ferromagnetic (spiral) configuration in the normal (superconducting) state. Minimizing Eq.~(\ref{eq:I}) as a function of $q$ yields $q \propto \sqrt{\Delta}$ with a constant of proportionality that increases as $\mu=-2$ is approached, consistent with Fig.~\ref{fig:q_as_Delta}.

Higher order terms in $J$ represent multiple-scattering processes which, in particular, lead to the formation of localized subgap YSR states around each adatom with energy $\varepsilon(J)$ \cite{yu,shiba,rusinov,balatsky,yao14} (e.g. for a parabolic band and delta-function magnetic potential one finds $\varepsilon=\pm\Delta\frac{1-(\pi J\nu_F/2)^2}{1+(\pi J\nu_F/2)^2}$ where $\nu_F$ is the normal-state density of states at the Fermi level). Heuristically, one can understand the role the YSR states play in modifying the adatom magnetic order by appealing to the general results found for the case of a 3D substrate \cite{schecter16}. In particular, it was shown that the overlap of the YSR states can reinforce the spiral formation, due to the renormalization of the antiferromagnetic exchange term that arises from superconductivity \cite{yao14}. Essentially, the hybridization of a pair of YSR states with a Cooper pair in the substrate leads to an enhancement of the second term in Eq.~(\ref{eq:I}), which amounts to replacing the prefactor $\Delta$ by $\Delta^2/|\varepsilon|$. As a result, the wavevector increases as $q\sim \sqrt{\Delta(\Delta/|\varepsilon|)}$ and is thus enhanced by the factor $\sqrt{\Delta/|\varepsilon|}>1$. In the limit of a large substrate coherence length $\xi\gg a_{\rm ad}$, this scaling of $q$ is applicable for $|\varepsilon|>\Delta/\sqrt{k_Fa_{\rm ad}}$; for smaller values of $|\varepsilon|$ the YSR band (of width $\propto \Delta/\sqrt{k_Fa_{\rm ad}}$) crosses the Fermi level where ferromagnetic YSR double exchange occurs and favors a smaller value of $q$ \cite{schecter16}. The double exchange mechanism, discussed more in Sec.~\ref{sec:chain}, is controlled by the kinetic energy of the YSR band and is not captured by the effective Heisenberg model Eq.~(\ref{eq:Heis}).

\begin{figure*}[t!]
\centering
\includegraphics[width=\textwidth]{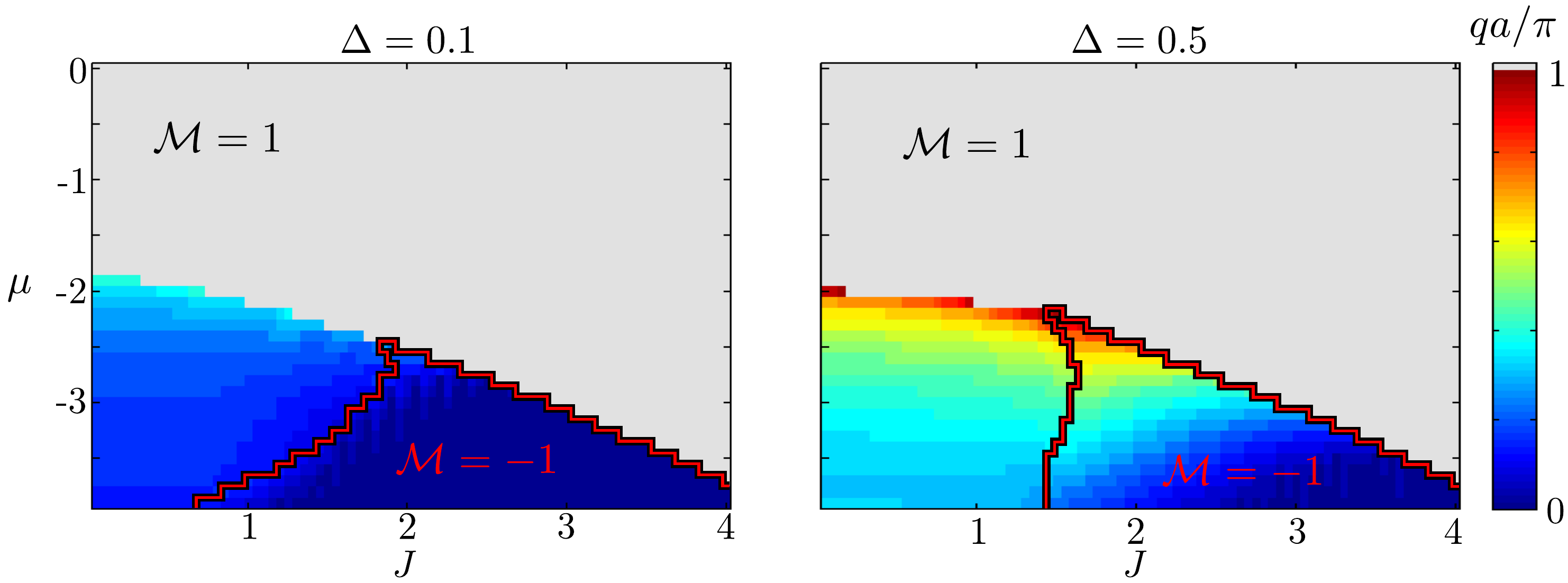}
\caption{\label{fig:phase_diagrams} (Color online) Phase diagram for a chain along (10) with $\Delta=0.1$ (left) and $\Delta=0.5$ (right), and $N_{x}\times N_{y}=120\times 71$. Grey (blue) denotes an antiferromagnetic (ferromagnetic) state. The red lines denote the border between domains of Majorana number $\mathcal{M}=1$ (trivial) and $\mathcal{M}=-1$ (non-trivial). For low fillings and small $J$ the behaviour depicted is consistent with the expectation that superconductivity aids in the formation of a spiral phase. The transition between spiral and antiferromagnetic phases is first order.}
\end{figure*}

We thus find that the spiral wavevector exhibits a small peak as a function of $\varepsilon$ (or $J$) near the topological superconducting transition of the order $q_{\rm max}\sim (k_Fa_{\rm ad})^{1/4}q(J\to0)$, i.e. there is a weak relative enhancement of $q$ proportional to $(k_Fa_{\rm ad})^{1/4}$ compared to $q$ in the small $J$ limit. Consequently, for a 2D substrate, $q$ depends very weakly on small to moderate exchange couplings, and only deviates substantially from the $q(J\to0)$ value when the YSR band crosses the Fermi level and activates the ferromagnetic double exchange. Thus, in contrast to the case of a 3D substrate (where the dependence of $q$ on $\varepsilon,\,J$ is much stronger \cite{schecter16}), for a 2D substrate the magnetic order of the adatom chain at the topological transition can be understood rather well simply by studying the adatom magnetic susceptibility for weak exchange coupling, as shown in Figs.~\ref{fig:I_of_q}, \ref{fig:q_as_Delta}. For chains along the (10) direction this  conclusion is consistent with the numerical data presented in Sec.~\ref{sec:chain} even for the case $k_Fa_{\rm ad}< 1$, and for chemical potentials away from the band bottom (cf. Figs.~\ref{fig:phase_diagrams} and \ref{fig:10_vs_11}).

\subsection{Exchange interaction along the (11) direction}\label{sec:exchange_11}

Along the (11)-direction the exchange interaction behaves quite differently when the substrate is near half-filling, $\mu=0$. This is because the Fermi surface contains segments along the diagonals with very little curvature in the $(k_x, k_y)$-plane as well as segments along the axes with large curvature, see Fig.~\ref{fig:system_config}. This implies that the Fermi surface is nested and the electron Green function has spectral weight focused along the (11) and (-11) directions in real space \cite{aristov1} and thus displays effectively 1D behavior along the adatom chain. As a result, one expects the adatom chain to exhibit a $2k_F$ singularity in susceptibility, leading to $q=2k_F$ spiral order even in the \emph{absence} of superconductivity \cite{klinovaja13,braunecker13,vazifeh13,schecter15}. Here $k_F$ is defined as the Fermi momentum along the chain direction (i.e., for a (11) chain, $k_F$ is taken along the diagonal $k_x=k_y$). The $q=2k_F$ spiral order, based on perturbation theory, should be valid away from points of commensurability between $2k_F$ and $\pi/a_{\rm ad}$ \cite{schecter15}. For small $|\mu|$ (where $2k_Fa_{\rm ad}\approx 2\pi$), this implies that perturbation theory is valid for $J\ll|\mu|$ (where $q\approx 2k_F$), while for $J\gtrsim|\mu|$ we expect the system to lock into the commensurate ferromagnetic state \cite{schecter15}. This is consistent with the numerical data presented in Sec.~\ref{sec:chain} for the adatom chain where the magnetic order is determined by minimizing the total energy, Eq.~(\ref{eq:fullh}), for large $J$ and $|\mu|$ (cf. Fig.~\ref{fig:10_vs_11}).
\begin{figure}
\centering
\includegraphics[width=0.9\columnwidth]{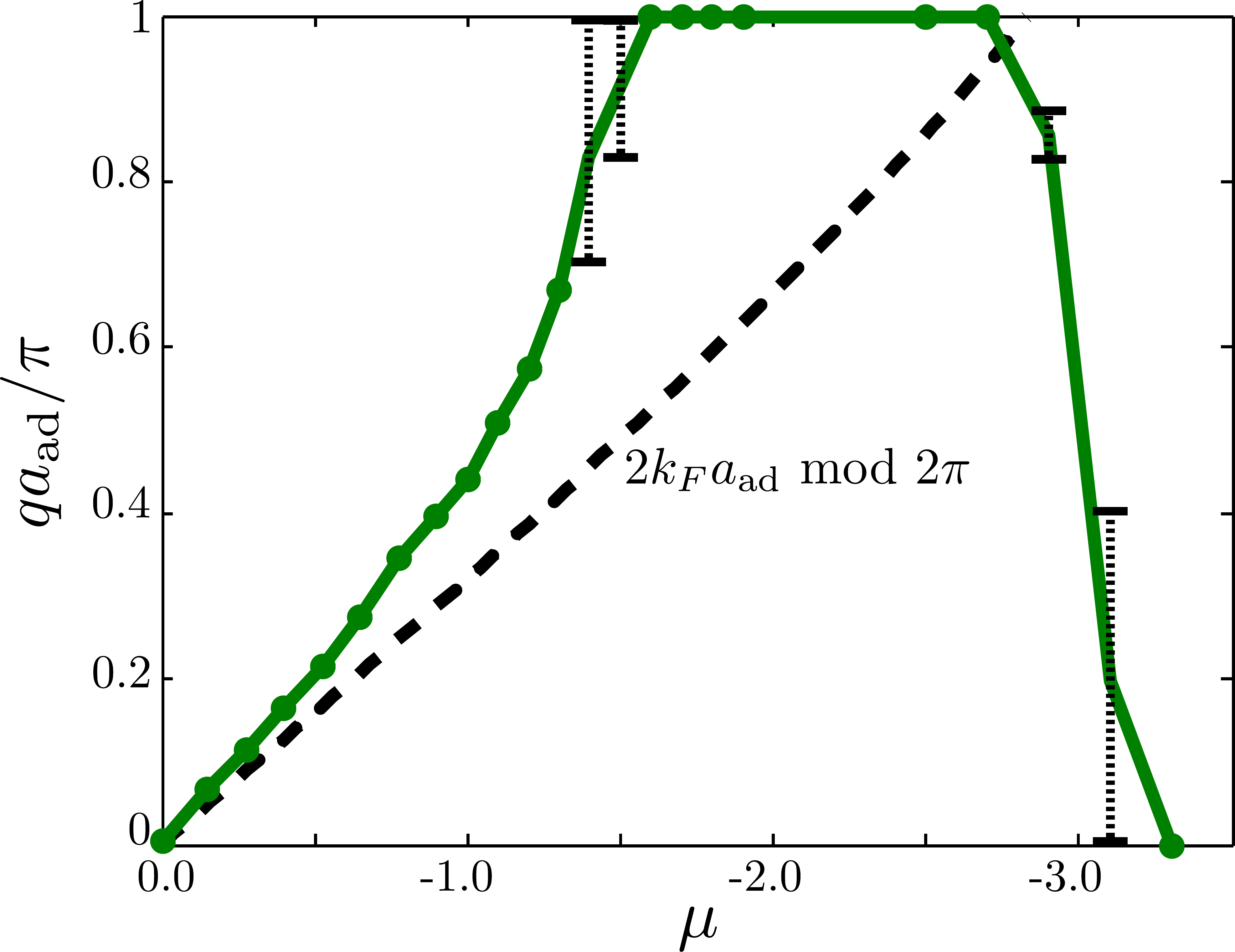}
\caption{\label{fig:q_as_mu_11} (Color online) Ground state wavevector $q$ (green curve) of an adatom chain along the (11) direction calculated from Eq.~(\ref{eq:fullh}) to leading order in $J$ as a function of $\mu$ for $\Delta=0$. The black dashed line is $2k_Fa_{\rm ad}$, where $k_F$ is defined as the Fermi momentum along the (11) direction $(k_x=k_y)$. The black dotted lines represent error bars inferred from the region of $\Omega(Q)$ which is essentially flat and therefore does not allow a reliable determination of the minimum.}
\end{figure}

We now verify the $q=2k_F$ behavior that exists for small $J$ by computing $q$ from Eq.~(\ref{eq:fullh}) to leading order in $J$ with adatoms placed along the (11)-direction, i.e. $a_{\rm ad}=\sqrt{2}a$.
In Fig.~\ref{fig:q_as_mu_11} the evolution of $q$ and $2k_F$ as a function of chemical potential is plotted and confirms the $q=2k_F$ behavior near $\mu=0$. The deviation of $q$ from $2k_F$ is expected as $|\mu|$ increases since the Fermi surface becomes more isotropic. Below a critical value of the chemical potential, $|\mu|\approx1.5$ a transition to an antiferromagnetic state occurs. We cannot determine within our resolution whether this transition is first or second order, as indicated by the error bars in Fig.~\ref{fig:q_as_mu_11}. As $|\mu|$ is increased further a second transition occurs to a spiral state that exists in the interval $2.6\lesssim|\mu|\lesssim 3.1$, before finally transitioning into a ferromagnet for larger $|\mu|$. According to Eq.~(\ref{eq:I}) there should be a transition between ferromagnetic and antiferromagnetic phases when $k_F a_{\rm ad} = \pi/2$ (corresponding to the integer $n=0$ above), or $|\mu|=2\sqrt{2}\approx 2.8$. This is roughly consistent with Fig.~\ref{fig:q_as_mu_11}, except that the first order antiferromagnet to ferromagnet transition at $|\mu|\approx 2.8$ is broadened into a narrow region of spiral order. Similar to the (10) direction, we find that antiferromagnetic order is stable against superconductivity, while ferromagnetic order is unstable to spiral formation with $q\propto \sqrt{\Delta}$. This is to be expected since ferromagnetic order in Fig.~\ref{fig:q_as_mu_11} occurs when the Fermi surface is approximately isotropic.

\section{Magnetic adatom chain}\label{sec:chain}

As a chain of impurities is formed, the YSR subgap states localized at the impurities hybridize and a band develops inside the superconducting gap. To account for the effects of this band, we go beyond the two-spin exchange approximation considered above, and numerically calculate from the total electronic energy, Eq.~(\ref{eq:therm-dyn}), the preferred magnetic order for a chain of magnetic adatoms within a coplanar variational ansatz
\begin{align}\label{eq:spin_vector_assumption}
	\mbf{S}_i = S\left(\phantom{A}^{\phantom{A}}\hspace*{-5mm}\cos(Q x_i),\,\sin(Q x_i),\, 0\hspace*{-5mm}\phantom{A}^{\phantom{A}}\right)
\end{align}
parametrized by the wavevector of the chain $Q=\frac{2\pi}{Na}$, where $N$ is an integer divisor of the number of adatom impurities. The choice of spin rotation axis as in Eq.~(\ref{eq:spin_vector_assumption}) can be made without loss of generality in the absence of SOC. Including the SOC shifts the value of $q$ but does not affect the topological phase boundaries, see Appendix~\ref{sec:spin_orbit}. Along the (10) direction we let the chain extend over the entire length (120 sites) of the system, and we impose periodic boundary conditions, as indicated in the left panel of Fig.~\ref{fig:system_config}(a). A system width of 71 sites along the (01) direction is used. For chains along (11) we employ a $71 \times 71$ system and place adatoms along the $x=y$ line. Periodic boundary conditions are imposed in this case as well, but contrary to the (10) case we carry out all calculations in real space, making them more demanding.

To facilitate efficient computations, we follow Ref.~\onlinecite{reis14} and perform a local spin-rotation,
\begin{eqnarray}\label{eq:gtransf}
	c_{i\sigma} \rightarrow {\tilde c}_{i\sigma}=e^{i\frac{\sigma}{2}qx_i}c_{i\sigma}\,,
\end{eqnarray}
which leaves $\mathcal{H}_{\text{SC}}$ invariant and transforms $\mathcal{H}_{0}$ and $\mathcal{H}_{imp}$ to
\begin{eqnarray}
\tilde{\mathcal{H}}_{0} &=& -\sum_{\substack{\langle ij \rangle \\ \alpha}} \tilde{t}_{ij,\alpha}\tilde{c}^{\dagger}_{i\alpha}\tilde{c}_{j\alpha} - \mu\sum_{i\sigma}\tilde{c}^{\dagger}_{i\alpha}\tilde{c}_{i\alpha} \,,\label{eq:h02}\\
\tilde{\mathcal{H}}_{\text{imp}} &=& J_{\text{imp}}S\sum_{\substack{i\in \mathcal{I} \\ \alpha \beta}}\tilde{c}^{\dagger}_{i\alpha}\sigma^{x}_{\alpha\beta}\tilde{c}_{i\beta}.
\end{eqnarray}
In the rotated basis the spin chain is a ferromagnet polarized along $\hat{x}$, while the hopping amplitude becomes spin- and wavevector-dependent
\begin{eqnarray}
	t\rightarrow \tilde{t}_{ij,\sigma}=te^{-i\frac{\sigma}{2}q(x_i - x_j)}\,,
\end{eqnarray}
where $x_i - x_j=\pm 1$ in units of the lattice constant. This transformation renders the Hamiltonian translationally invariant along the $x$-axis, and allows one to partially diagonalize the Hamiltonian using the Fourier transform
\begin{eqnarray}
	\tilde{c}_{i\alpha} = \sum_{k_{x}}e^{ik_{x}x_i} \tilde{c}_{k_{x}\alpha}(y_{i})\,,
\end{eqnarray}
with $k_{x}\in[-\pi/a,\pi/a[$. This reduces the time needed to obtain the full spectrum by a factor of $\sim N_x^2$. In the following, we evaluate the free energy of the system for 31 values of $q\in [0,\pi/a]$, which now enter exclusively via the hopping amplitudes $\tilde{t}_{ij,\sigma}$.

\subsection{Phase diagram}\label{sec:phase_diagram}

Here we consider the evolution of the magnetic order of the chain with changing chemical potential, adatom potential strength and superconducting order parameter. As is shown in Sec.~\ref{sec:selfconsistency}, including the feedback from the impurities on the local pairing potential in a selfconsistent manner does not significantly alter the magnetic or topological phases. Selfconsistency is therefore neglected in the remainder of this section. This also implies neglecting the other effect of selfconsistency, namely an overall suppression of the superconducting order parameter with changing chemical potential due to a reduction of the number of states available for pairing. The magnitude of the order parameter can thus be varied independently of the chemical potential. In Fig.~\ref{fig:phase_diagrams} we show phase diagrams corresponding to $\Delta=0.1$ and $\Delta=0.5$, which reveal behavior consistent with the general trends found in Ref.~\onlinecite{schecter16} when the Fermi surface is approximately isotropic, as discussed in Sec.~\ref{sec:exchange}.

\begin{figure}[t]
\centering
\includegraphics[width=\columnwidth]{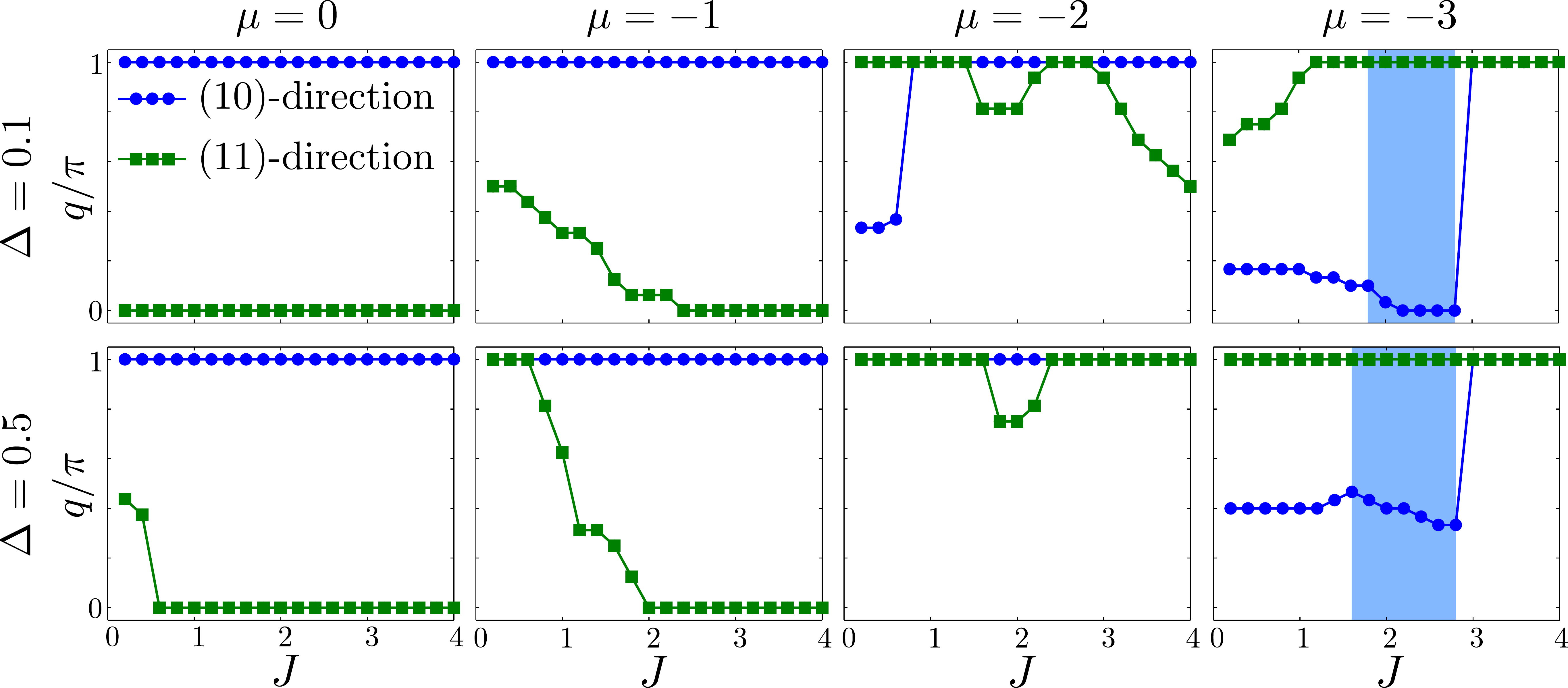}
\caption{\label{fig:10_vs_11} (Color online) Comparison of the ground state wavevector $q$ of an adatom chain along the (10) (blue) and (11) (green) directions. Data for the (10) chain correspond to cuts at fixed $\mu$ through the phase diagrams in Fig.~\ref{fig:phase_diagrams}. The light-blue shaded regions indicate the topologically non-trivial phase for the (10) direction, as well as the onset of YSR double exchange. Data for the (11) chain are consistent with the concept that the substrate behaves effectively as a 1D superconductor near $\mu=0$ (cf. Fig.~\ref{fig:q_as_mu_11} and the discussion in Sec.~\ref{sec:exchange_11}).}
\end{figure}

In particular, the analysis of Sec.~\ref{sec:exchange} predicts the magnetic order to be antiferromagnetic for $k_Fa_{\rm ad}>\pi/2$, where $k_F$ is the Fermi momentum along the chain direction. This translates to antiferromagnetic order for $0<|\mu|\lesssim2$ and spiral order for $|\mu|\gtrsim2$, which for small $J$ agrees well with the phase diagrams in Fig.~\ref{fig:phase_diagrams}  determined by minimizing $\Omega(Q)$. The superconductivity induced antiferromagnetic contribution to the exchange interaction, which is proportional to $\Delta$ [see Eq.~(\ref{eq:I})], slightly shifts the boundary between antiferromagnetic and spiral phases, thus accounting for the small difference between the $\Delta=0.1$ and $\Delta=0.5$ cases in Fig.~\ref{fig:phase_diagrams}. For $J \ll 1$ and a (10) chain, the magnetic order weakly depends on $J$ (see Figs.~\ref{fig:phase_diagrams} and \ref{fig:10_vs_11}), however, as $J$ is increased the YSR band eventually crosses the Fermi level. As mentioned in Sec.~\ref{sec:exchange}, this activates the ferromagnetic YSR double exchange mechanism \cite{schecter16} and leads to a decrease in the wavevector $q$ with increasing $J$. This behavior is shown in the last column of Fig.~\ref{fig:10_vs_11}.

The transition to an antiferromagnetic state at larger $J$ occurs in the absence of superconductivity. It is also reflected in the two-spin exchange coupling, indicating that it is not a multi-spin effect. Therefore, one could capture this effect by mapping the evolution of $q$ as a function of $\mu$ and $J$  including higher-order corrections in $J$ to the two-spin exchange interaction, Eq.~(\ref{eq:I}). In our model, the decrease of the antiferromagnetic phase boundary line occurs already at quartic order in $J$, but whether this particular behavior is generic remains an open problem.

\begin{figure*}
\centering
\includegraphics[width=\textwidth]{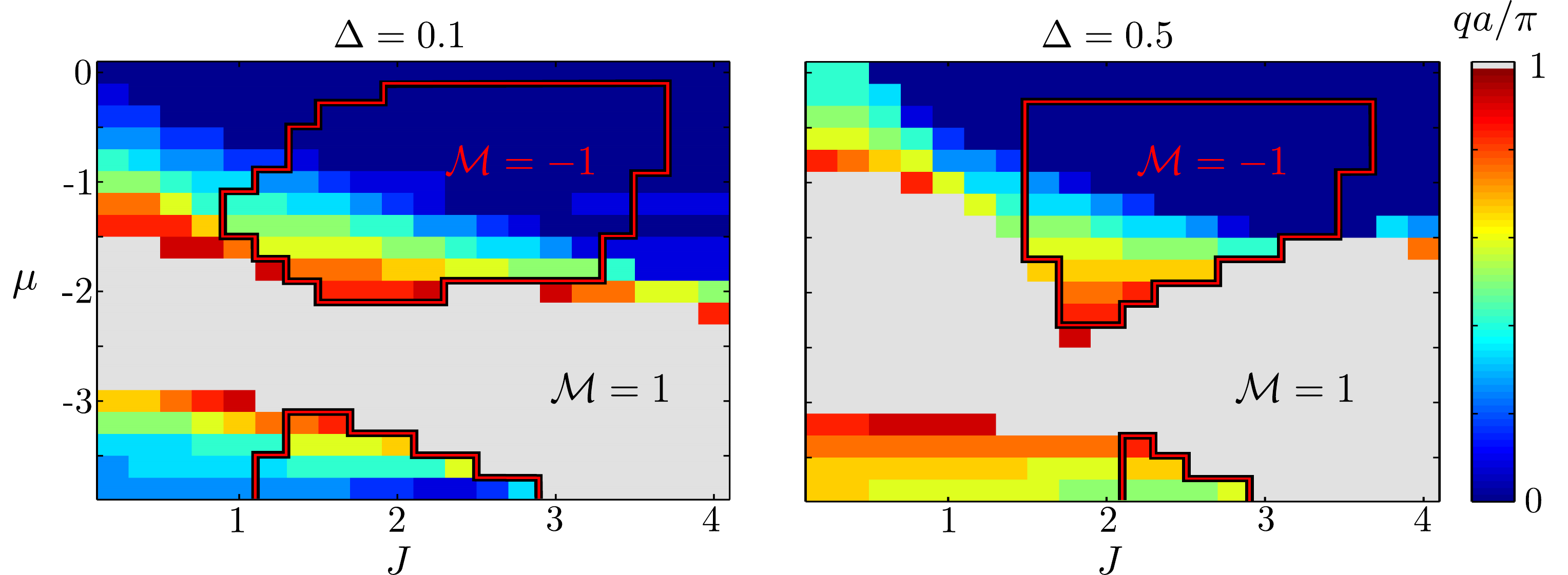}
\caption{\label{fig:phase_diagrams_11} Phase diagrams for a chain along (11) with $\Delta=0.1$ (left) and $\Delta=0.5$ (right). Here $N_x \times N_y = 48 \times 48$. For $|\mu| \lesssim 2$ the behavior is found to match expectations from 1D, where a spiral yields to a ferromagnet when $J \sim |\mu|$. 2D behavior is recovered for $|\mu| \gtrsim 2$, although recall that the adatom spacing is modified. The red lines denote the border between domains of Majorana number $\mathcal{M}=1$ (trivial) and $\mathcal{M}=-1$ (non-trivial). The resolution is different from Fig.~\ref{fig:phase_diagrams} as the determination of $q$ is substantially more demanding in real space.}
\end{figure*}

As discussed in Sec.~\ref{sec:exchange_11}, there can be substantial differences between forming the adatom chain along the (10) and (11) crystallographic directions. In addition to a modification of the adatom spacing, the (11) direction also nests the Fermi surface near half-filling and this leads to the possibility of spiral order in the absence of superconductivity. In Fig.~\ref{fig:phase_diagrams_11} we plot $q$ for $\Delta=0.1$ and $\Delta=0.5$ for a chain along (11). Behavior distinct from the (10) direction is evident in particular for $|\mu| \lesssim 2$ where the Fermi surface nesting is the most prominent. For $|\mu|\approx 3$ the spiral phase appears and yields to an antiferromagnetic phase as $J$ is increased. At this point the Fermi surface is nearly isotropic and the system exhibits behavior similar to the (10) direction with a slightly larger adatom spacing compared to the case considered above. To highlight the differences between (10) and (11) we plot $q$ in both cases as a function of $J$ for cuts at fixed values of $\mu$ in Fig.~\ref{fig:10_vs_11}. For $\mu=-1,\,\Delta=0.1$, where Fermi surface nesting is still active, one finds the wavevector for $J\to0$ in Fig.~\ref{fig:10_vs_11} to differ only slightly from the value $qa_{\rm ad}/\pi\approx0.45$ shown in Fig.~\ref{fig:q_as_mu_11} (the discrepancy is due to finite $\Delta=0.1$ in the case of the former). As $J$ is increased, however, $q$ rapidly decreases until $J\approx 2$, beyond which it saturates. This is consistent with the result of Ref.~\cite{schecter15,hu15} for a 1D substrate that predicts a second order transition from a spiral into a ferromagnetic state at a critical value of $J$ proportional to the deviation from commensurability. For a chain along the (11) direction this would occur for $J\sim|\mu|$, which appears to be consistent with Fig.~\ref{fig:phase_diagrams_11} and Fig.~\ref{fig:10_vs_11}. Contrary to a chain along (10), the ferromagnetic state along (11) for $\mu=0$ is more robust towards the addition of superconductivity. We found a ferromagnetic ground state for systems up to $100 \times 100$. This suggests that if spiral order occurs for larger systems, the value of $qa$ is smaller than $\pi / 50 $.

\subsection{Topological phases}

By evaluating the Majorana number we can distinguish phases of trivial and non-trivial topology. The Majorana number is defined as~\cite{kitaev01}
\begin{eqnarray}
	\mathcal{M}=\text{sign} \left( \text{Pf}[\mathcal{A}(0)] \text{Pf} [\mathcal{A}(\pi)] \right)\,,
\end{eqnarray}
where $\widetilde{H}(k)=\tfrac{i}{4}\mathcal{A}(k)$ is the Hamiltonian in the Majorana representation and Pf denotes the Pfaffian. In Fig.~\ref{fig:phase_diagrams} regions with negative Majorana number, denoting the non-trivial phase, are bounded by red lines. We remind the reader that this is not a sufficient condition for the phase to support localized Majorana modes, as there should also be a quasiparticle gap, i.e. the Majorana modes should be separated from the bulk YSR spectrum by an energy gap. The topological gap depends sensitively on the magnetic order of the adatom chain \cite{pientka1}. As $q$ decreases and the magnetic order approaches ferromagnetism, the topological gap decreases and is strictly zero for $q=0$. This is because singlet Cooper pairs in the substrate cannot tunnel into a spin-polarized YSR chain. We generally find that the presence of strictly ferromagnetic, $q=0$, configurations in Fig.~\ref{fig:phase_diagrams} appear to be a consequence of finite size effects, which quantize the value of $q$ under periodic boundary conditions. We have confirmed that with increasing system size (to $N_x \times N_y = 240 \times 101$) the ferromagnetic phase for a (10) chain indeed becomes a weak spiral. 

\begin{figure}
\centering
\includegraphics[width=0.9\columnwidth]{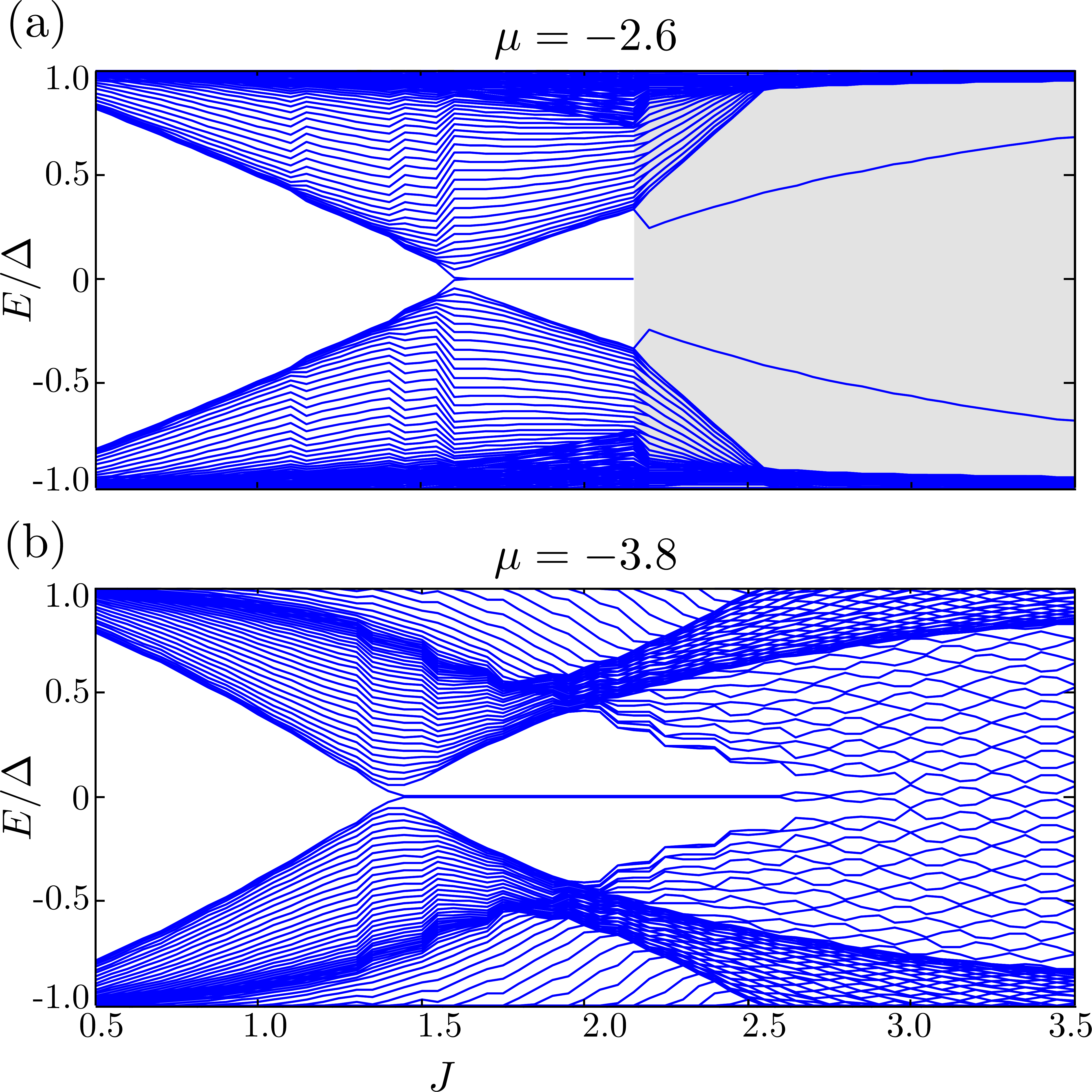}
\caption{\label{fig:spaghetti_plots_two_mu} (Color online) Electron energy spectrum as a function of $J$ for a (10) chain and two values of $\mu$ (here $\Delta=0.5$). (a) The topological phase terminates at large $J$ due to a first order magnetic transition into an antiferromagnetic state, indicated by the grey region. Two fermionic subgap states persist in the antiferromagnetic region and are localized to the chain boundaries. (b) The Majorana modes delocalize and hybridize when the topological gap closes as a result of ferromagnetic order $q=0$, however ferromagnetic order appears to be a finite size effect.}
\end{figure}

In Fig.~\ref{fig:spaghetti_plots_two_mu}(a) we plot the electron energy spectrum as a function of $J$ for $\mu=-2.6,\,\Delta=0.5$, showing the energy gap closing and reopening across the topological transition. Within the non-trivial phase there exists a pair of states near zero energy, which indicate the presence of Majorana bound states weakly hybridized due to the finite extent of the chain. A first order transition to the antiferromagnetic phase occurs at larger $J$ (indicated by the grey region) and coincides with the abrupt termination of the zero energy state, see Fig.~\ref{fig:spaghetti_plots_two_mu}(a). 
This differs substantially from the case when the topological gap closes due to the formation of a ferromagnetic state, see Fig.~\ref{fig:spaghetti_plots_two_mu}(b). We note that the closing of the topological gap for $J\gtrsim 2$ in Fig.~\ref{fig:spaghetti_plots_two_mu}(b) reflects the decrease of $q$ with $J$ in Fig.~\ref{fig:phase_diagrams}.

The remaining subgap states seen in Fig.~\ref{fig:spaghetti_plots_two_mu} in the antiferromagnetic phase can be understood in terms of an effective two-channel $p$-wave superconductor, where each channel supports a Majorana bound state at each end of the chain. The hard-wall boundary condition hybridizes these states to create a single localized fermionic state at each end of the chain~\cite{pientka1}. 

\begin{figure}
\centering
\includegraphics[width=\columnwidth]{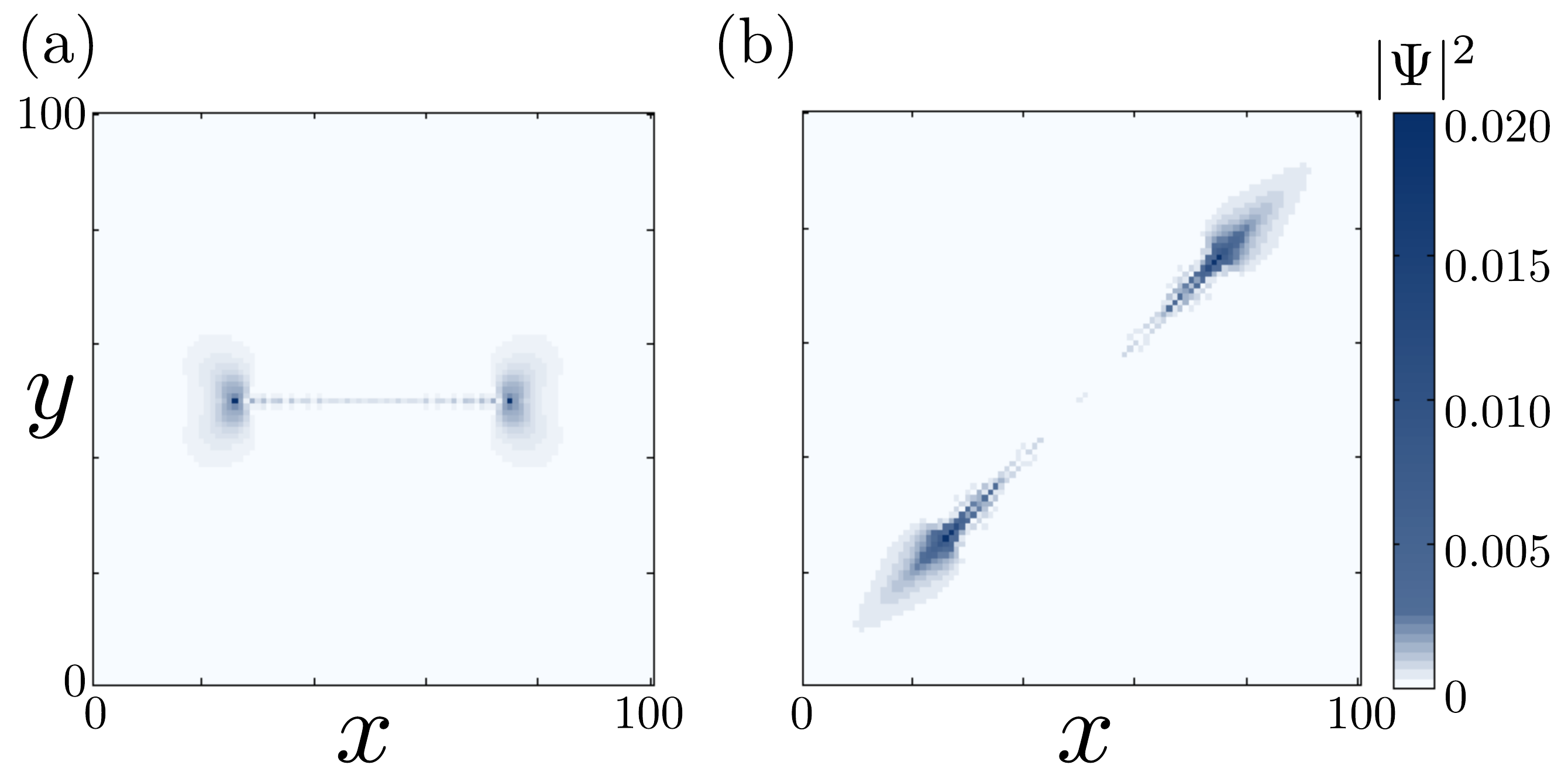}
\caption{\label{fig:majorana_end_modes} (Color online) (a) Majorana end mode for a chain along (10) for $\mu=-2.6$ and $J=2.1$. (b) Majorana end mode for a chain along (11) for $\mu=-0.65$ and $J=1.8$. In both cases $\Delta=0.1$. The localization length along the chain depends sensitively on the chosen parameters.}
\end{figure}

The (11)-direction also exhibits Majorana bound states, albeit for different parameter values, consistent with the fact that the chain is parallel to the nesting wavevector and has a larger lattice spacing. Thus, the topologically non-trivial region already occurs for $\mu$ close to half filling (but not for $\mu=0$), and for $J \approx 1$. In Fig.~\ref{fig:majorana_end_modes} the bound states along the two different directions are illustrated.

\subsection{Effect of direct exchange interaction}

\begin{figure*}
\centering
\includegraphics[width=\textwidth]{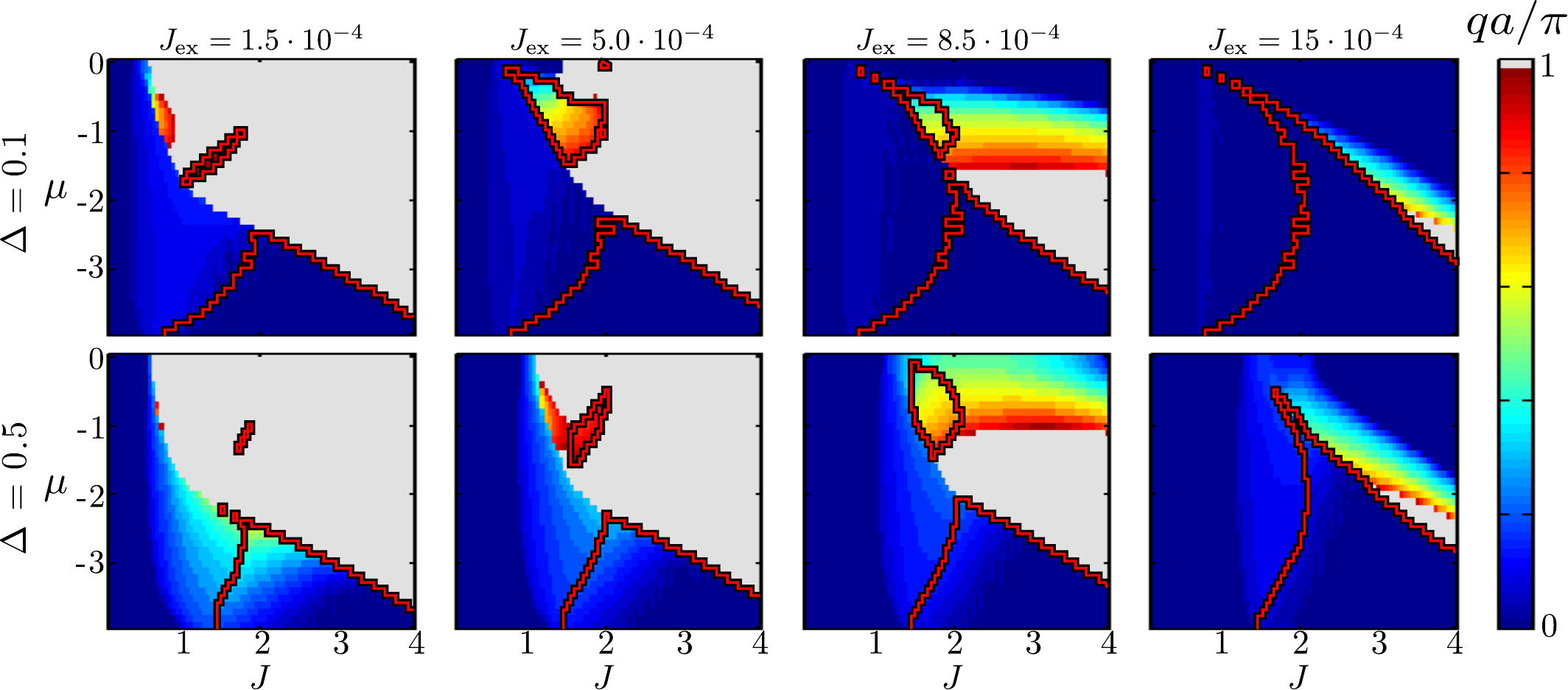}
\caption{\label{fig:direct_exchange} (Color online) Phase diagrams illustrating the effect of adding a ferromagnetic direct exchange term between the adatomic impurities. The red outline denotes the boundaries between regions of $\mathcal{M}=1$ and $\mathcal{M}=-1$.}
\end{figure*}

Motivated by the close proximity of the adatoms, we briefly remark on the consequences of having an additional direct, nearest neighbor ferromagnetic exchange interaction between them. We assume the adatoms to lie along the (10) direction with $a_{\rm ad}=a$ and minimize the total energy
\begin{eqnarray}
	E_{\text{tot}}(Q) = \tilde{\Omega}(Q) - J_{\text{ex}} \cos Qa_{\rm ad}\,,
\end{eqnarray}
where  $J_{\text{ex}}>0$ denotes the strength of the direct exchange interaction and $\tilde{\Omega}=\Omega/N_{\rm ad}$ is the thermodynamic potential per adatom. In Fig.~\ref{fig:direct_exchange} we show how the direct exchange modifies the phase diagram for increasing values of $J_{\text{ex}}$. We find that as $J_{\text{ex}}$ is increased the antiferromagnet/spiral phase boundary line shifts to make the antiferromagnetic region smaller, and the ferromagnetic or weak spiral phases larger. At the same time, new regions of strong spiral order with $qa/\pi\sim0.5$ open near half-filling, previously occupied by the antiferromagnetic phase. This occurs for a moderate exchange coupling $J_{\rm ex}\sim 5 \cdot10^{-4}$, which is roughly 1/4 of the scale set by the indirect exchange coupling in that region of parameters. The latter can be estimated, e.g., by calculating the magnetic energy bandwidth near $\mu=-1,\,J=1.5$ (cf. Fig.~\ref{fig:direct_exchange}), defined as the difference between the maximum and the minimum of $\tilde{\Omega}(Q)$.

Another interesting feature to observe is that the spiral phase can survive in the presence of rather large $J_{\rm ex}$. In the case with $\Delta=0.1$ the spiral phase remains for $J_{\text{ex}}\lesssim 15 \cdot 10^{-4}$ in a narrow vertical region near $J\approx 0.7$ in Fig.~\ref{fig:direct_exchange}. For $\Delta=0.5$ the spiral phase can be found for $J_{\text{ex}}\lesssim 8 \cdot 10^{-3}$ in a wider vertical region near $J=1.5$. For $\Delta=0.1$, the maximal value of $J_{\rm ex}$ exceeds the indirect exchange coupling (evaluated in the narrow region where the spiral last existed) by a factor of 8-10, while for $\Delta=0.5$ the maximal value of $J_{\rm ex}$ is 4 times larger than the indirect exchange. 

The robustness of spiral order with respect to such large values of the direct exchange interaction can be traced back to the long-range nature of the indirect antiferromagnetic exchange coupling in Eq.~\ref{eq:I}. Adding a direct exchange interaction leads to a total energy that may be expressed for $\xi^{-1}\ll Q\ll\pi/a_{\rm ad}$ as 
\begin{align}\label{eq:I_direct}
E_{\rm tot}(Q)=\frac{1}{2}(J_{\rm ex}+J_{\rm RKKY})(Qa_{\rm ad})^2-J_{\rm RKKY}\frac{a_{\rm ad}}{\xi}\mathrm{ln}(Qa_{\rm ad})
\end{align}
where for $J\sim 1$ we have $J_{\rm RKKY}\sim v_F/(k_Fa_{\rm ad}^2)$ (cf. Sec.~\ref{sec:exchange}). Minimizing $E_{\rm tot}$ leads to 
\begin{align}\label{eq:q_direct}
qa_{\rm ad}=\sqrt{\frac{a_{\rm ad}}{\xi}\frac{J_{\rm RKKY}}{J_{\rm ex}+J_{\rm RKKY}}}.
\end{align}
The expression in Eq.~(\ref{eq:q_direct}) holds only for $q>\xi^{-1}$, or $J_{\rm ex}<J_{\rm RKKY}(\xi/a_{\rm ad}-1)$, while for larger $J_{\rm ex}$ the true ground state is a ferromagnet. If the chemical potential lies within the YSR band, an exponentially small $q$, with exponent proportional to $J_{\rm ex}/\Delta$, is expected due to a gain in YSR condensation energy~\cite{schecter16}. For the topologically trivial regime we find for $\xi/a_{\rm ad}\gg 1$ that a spiral phase exists even for parametrically large $J_{\rm ex}$,
\begin{align}\label{eq:window}
	J_{\rm ex}<J_{\rm RKKY}(\xi/a_{\rm ad})
\end{align}
implying that a window exists in which the direct exchange interaction exceeds the indirect RKKY exchange interaction but a spiral phase still occurs. The existence of this window ultimately stems from the scaling law $q\propto\sqrt{\Delta}$ discussed in Sec.~\ref{sec:exchange}. For a 3D substrate one has $q\propto\Delta$ and the window in Eq.~(\ref{eq:window}) is absent (i.e. the adatom chain becomes ferromagnetic once $J_{\rm ex}\gtrsim J_{\rm RKKY}$). We also note that although the window becomes larger with increasing $\xi$ (decreasing $\Delta$), it also has the adverse effect of decreasing the magnitude of $q$, see Eq.~(\ref{eq:q_direct}). These considerations appear qualitatively consistent with the numerical data shown in Fig.~\ref{fig:direct_exchange} and discussed above.

\section{Effects of selfconsistency}\label{sec:selfconsistency}

\begin{figure}[!t]
\centering
\includegraphics[width=\columnwidth]{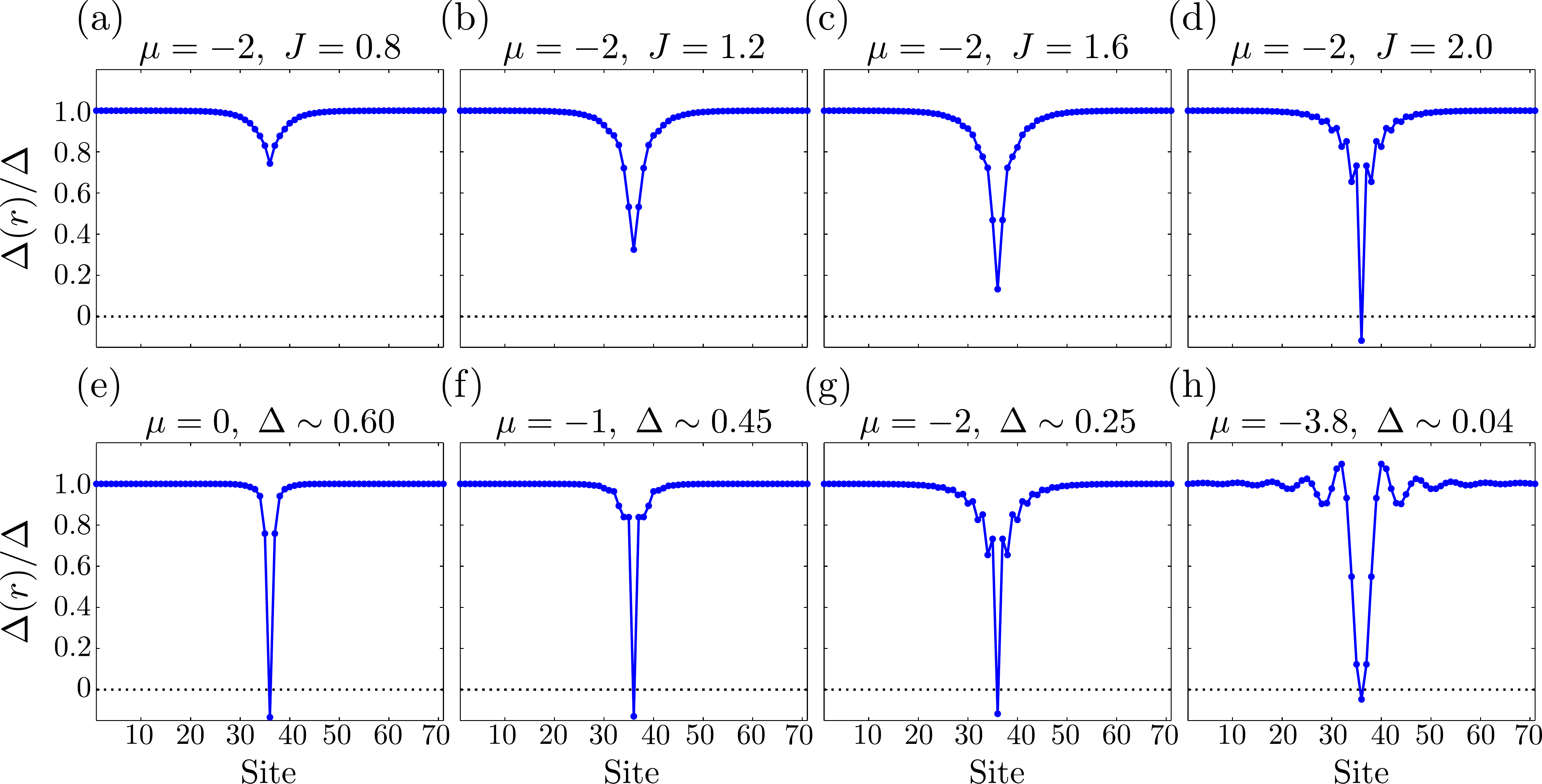}
\caption{\label{fig:local_SC_OP_suppression} (Color online) Effect of self-consistency on the superconducting order parameter for $V=2.5$ for a chain along (10). (a)--(d) Evolution of the superconducting order parameter as $J$ is increased for a fixed value of the chemical potential. As $J$ is increased above $J_c$, the order parameter along the chain changes sign. (e)--(h) Superconducting order parameter for different values of $\mu$ for fixed $J=2$. The extent of the suppression is seen to vary with the chemical potential~\cite{flatte97}.}
\end{figure}

Within selfconsistent mean field theory, the pair-breaking magnetic adatoms will give rise to a local suppression of the superconducting pair potential near the adatom chain~\cite{flatte97,balatsky,meng15}. This is illustrated in Fig.~\ref{fig:local_SC_OP_suppression}, where we plot the spatial profile of the pair potential across the width of the system, with the adatom chain along the (10) direction located on site number 36 along the (01) direction. For fixed chemical potential, the suppression is seen to increase with $J$, and even lead to an on-chain negative pair potential at $J=2$. For fixed $J=2$, on the other hand, the spatial modulation of the pair potential is seen to extend further from the chain when $\mu$ is lowered and the Fermi wavelength increases. Both of these trends are consistent with expectations based on results of Refs.~\onlinecite{flatte97,balatsky}.

\begin{figure}[!t]
\centering
\includegraphics[width=\columnwidth]{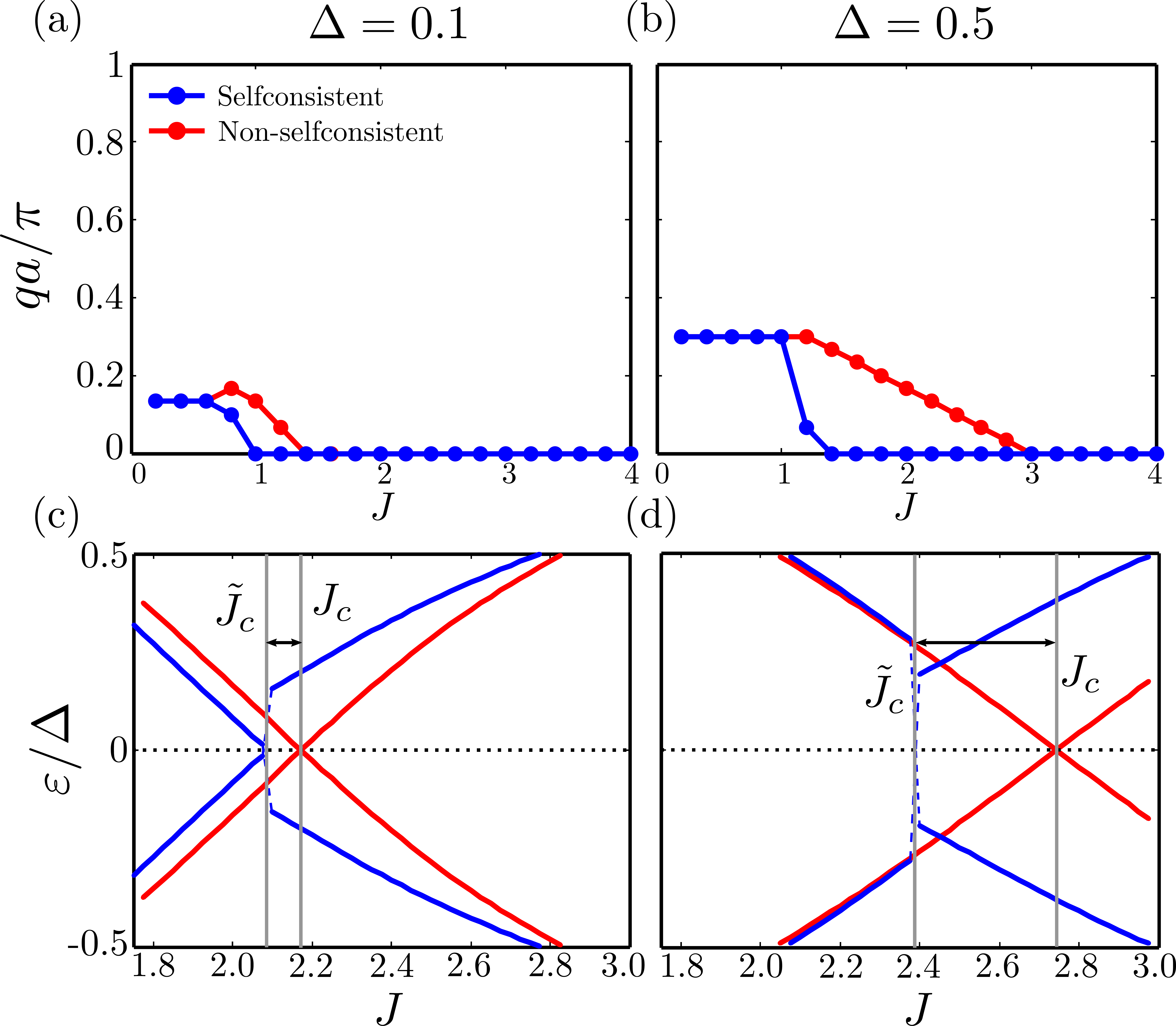}
\caption{\label{fig:selfcons_vs_nonselfcons} (Color online) (a)--(b) Comparison of selfconsistent and non-selfconsistent approaches when evaluating the minimum of the thermodynamic potential for an adatom chain. The selfconsistent approach accounts for the local suppression of the order parameter depicted in Fig.~\ref{fig:local_SC_OP_suppression}. Both cases depicted are for $\mu=-3.8$. Here $\Delta$ refers to pairing potential far away from the chain, $\Delta=0.1$ requires $V=3.09$ while $\Delta=0.5$ implies $V=4.80$. The effect depends on the magnitude of $\Delta$ as the suppression of the effective chemical potential felt by the subgap YSR states is smaller for smaller $\Delta$. (c)--(d)  Illustration of the reduction of $J_c$ for a single adatom by selfconsistency. The dependence on the magnitude of $\Delta$ is evident, the reduction in (d) is much more pronounced than in (c).}
\end{figure}

To determine to what extent the local suppression affects the magnetic order along the chain we compare the $q$-vector for selfconsistent and non-selfconsistent evaluations of the thermodynamic potential in Fig.~\ref{fig:selfcons_vs_nonselfcons}(a)-(b). The effect of selfconsistency is seen to be minor and dependent on the magnitude of the bulk gap $\Delta$, which is defined here as $\Delta_i$ evaluated far from, or in the absence of, the adatom chain. The effect of local suppression of the pairing potential can be understood as follows: The local pairing potential on the chain is suppressed leading to a decrease of the effective chemical potential, $\varepsilon(J)$, for the subgap YSR states~\cite{pientka1}. For a single adatom the YSR state crosses zero energy at $J=J_c$. Selfconsistency effectively reduces $J_c$ to $\tilde{J}_c$, which, for an adatom chain, causes the YSR band to cross the Fermi level at a smaller value of $J$ as depicted in Fig.~\ref{fig:selfcons_vs_nonselfcons}(c)-(d). Hence, ferromagnetic double exchange sets in at a lower $J$, leading to a reduction of $q$. This behavior is evident in Fig.~\ref{fig:selfcons_vs_nonselfcons}. We note that, as before, the appearance of a strictly ferromagnetic state in the selfconsistent calculation is a consequence of finite-size effects and increasing system size reveals a weak spiral state. As $\Delta$ is reduced, the renormalization of the effective chemical potential of the YSR band is reduced as well, thus making the effect less prominent. This is indicated in Fig.~\ref{fig:selfcons_vs_nonselfcons} and consistent with Ref.~\onlinecite{flatte97}.

Selfconsistency does not alter the phase boundary between spiral and antiferromagnetic states within the step-size used for $J$ ($=0.1$). The onset of antiferromagnetic order occurs even in the absence of superconductivity and can be understood by the higher-order corrections to the exchange interaction, as explained in Sec.~\ref{sec:phase_diagram}. This does not depend on the details of the gap and hence selfconsistency does not significantly shift the onset of antiferromagnetism. This is confirmed by comparing selfconsistent and non-selfconsistent calculations at a higher value of $\mu$ (not shown).
These observations allow us to disentangle the effects of superconductivity from those arising solely from varying the chemical potential when considering the magnetic order along the chain, as was done in Sec.~\ref{sec:chain}.

\begin{figure}
\centering
\includegraphics[width=0.9\columnwidth]{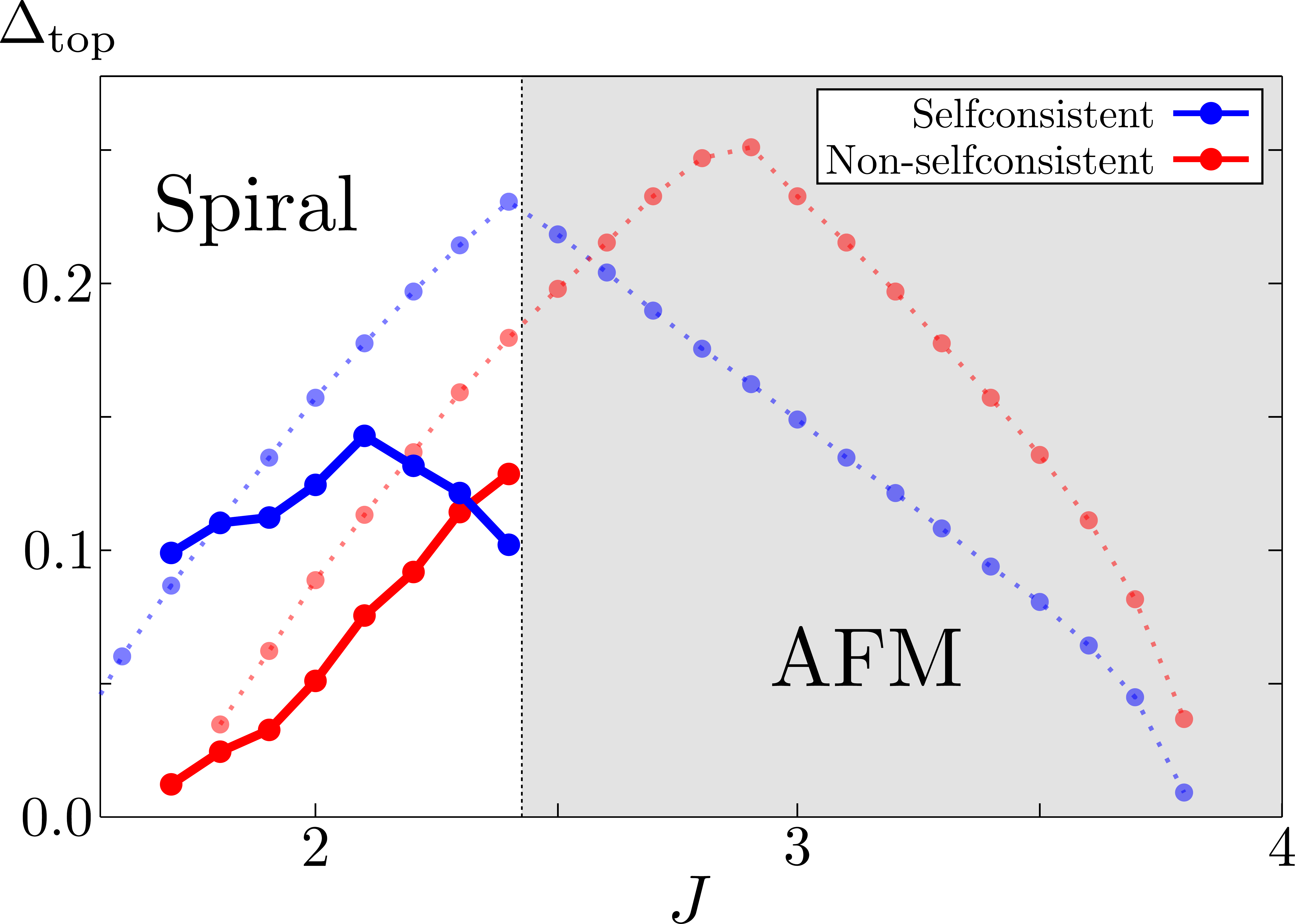}
\caption{\label{fig:topological_gap_vs_J} (Color online) Illustration of the shift of the maximum of the topological gap towards lower $J$ as the local pairing potential is suppressed by selfconsistency, while the bulk gap is kept fixed at $\Delta=0.5$. The full curves correspond to the case where the $q$-vector is determined by minimizing the energy (see Fig.~\ref{fig:phase_diagrams}). To illustrate the entire evolution of the dome, the faded curves were computed for a fixed $q$-vector ($qa=\tfrac{14\pi}{30}$). The faded dotted red curve has $\Delta=0.5$ on all sites while for the faded dotted blue curve, the pairing potential was suppressed locally at the chain sites to $\Delta(r_{i \in \mathcal{I}})=0.2$. In these plots $\mu=-2.8$.}
\end{figure}

Due to the reduction of $J_c$ by the local suppression of the pairing potential, selfconsistency also has an effect on the topological gap, as depicted in Fig.~\ref{fig:topological_gap_vs_J}. The topological gap exhibits non-monotonic behavior as a function of $J$, increasing from zero at the topological phase transition to a maximum at $J\sim J_c$ before decreasing to zero as antiferromagnetic order sets in and the topological phase ceases to exist. As above, the effect of selfconsistency is proportional to the magnitude of $\Delta$, and for $\Delta=0.1$ (not shown) the two cases are barely distinguishable.
Together with the Fermi velocity of the YSR band, $v_{F}^{\ast}$, the topological gap controls the Majorana localization length $\ell \sim v_{F}^{\ast} / \Delta_{\text{top}}$. However, $\ell$ appears to depend sensitively on parameters despite the fact that the topological gap exhibits the simple shape shown in Fig.~\ref{fig:topological_gap_vs_J}, which could be explained by a sensitivity to parameters in $v_F^{\ast}$.

\section{Conclusions}\label{sec:conclusions}

In this paper we performed a detailed study of the indirect exchange interactions between impurities deposited on two-dimensional superconducting substrates. We showed that spiral order can form along a chain of adatoms due to such interactions. One component of these is antiferromagnetic and owes its origin to the presence of superconductivity, while the other is the standard oscillating RKKY component. Unless the chain nests the Fermi surface the spiral order does not arise from a $2k_F$ peak in the susceptibility but instead from the competition between the superconducting antiferromagnetic component and the oscillating RKKY component. For a chemical potential near the band bottom we found the dependence $q\propto\sqrt{\Delta}$, implying a pronounced effect of superconductivity on the spiral $q$-vector. The exchange interactions along (11) for a system close to half filling were shown to exhibit behavior consistent with a chain of adatoms deposited on a 1D conductor; spiral order with $q \propto 2k_F$ forms for $|\mu| < 1$ as seen in Fig.~\ref{fig:q_as_mu_11}. This is in stark contrast to the exchange interactions along (10) which display antiferromagnetic behavior close to half filling. 

For a chain of impurities we contrasted selfconsistent and non-selfconsistent approaches and found that the local suppression of the pairing potential induced from the feedback of the impurities on the superconducting order parameter only affects the magnetic order around $J \sim J_c$ where the YSR band crosses the Fermi level. This allows us to study the phase diagram of the chain (in Fig.~\ref{fig:phase_diagrams}) without imposing selfconsistency and thus decoupling the chemical potential from the superconducting order parameter. For $J\lesssim J_c$, when the chemical potential lies outside the YSR band, the magnetic order is described by two-spin exchange interactions. The validity of the weak-coupling description is a consequence of the relatively weak dependence of the magnetic ordering vector $q$ on the YSR energy $\varepsilon$ (see Sec.~\ref{sec:exchange_10}). As the YSR band crosses the Fermi level, the exchange picture breaks down however, and $q$ is reduced by ferromagnetic double exchange. Including a direct exchange coupling between the adatoms allow the formation of spiral phases even for $J_{{\rm ex}} > J_{{\rm RKKY}}$ due to the strong dependence $q \propto \sqrt{\Delta}$ behavior found for a 2D substrate.

Topologically non-trivial regions of the phase diagrams are found in the spiral phases and exhibit Majorana bound states. The topological transition occurs as the YSR band crosses the Fermi level, at which point double exchange becomes a factor and $q$ is suppressed, see Fig.~\ref{fig:10_vs_11}. The topological gap in the non-trivial regions were found to exhibit a weak dependence on selfconsistency through the reduction of $J_c$ by the suppression of the local pairing potential.

\begin{acknowledgements}

The authors gratefully acknowledge P. Kotetes, M. Schulz, P. Orth, A. Kreisel, D. Scherer, and A. Black-Schaffer for helpful discussions. M.H.C. and B.M.A. acknowledge support from a Lundbeckfond Fellowship (Grant No. A9318). The Center for Quantum Devices is funded by the Danish National Research Foundation.

\end{acknowledgements}

\appendix

\section{Effect of spin-orbit coupling}\label{sec:spin_orbit}

\begin{figure}
\centering
\includegraphics[width=0.75\columnwidth]{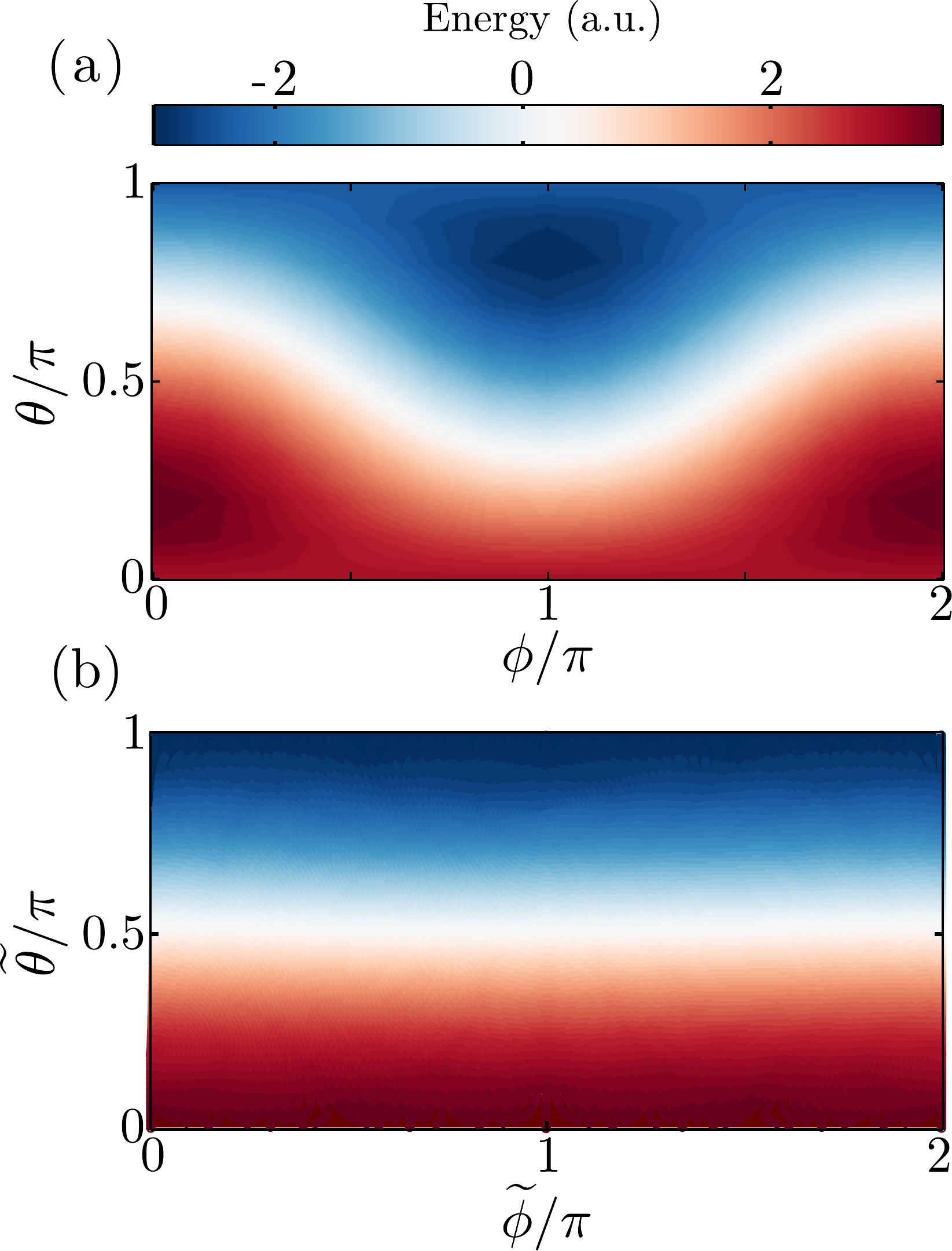}
\caption{\label{fig:tso_landscape} (Color online) Energy as function of both polar and azimuthal angles in the presence of spin-orbit coupling with $t_{\text{so}}=0.1$ and $J=0.1$ and the distance between the two adatoms $a_{\rm ad}=5$. In (a) the energy landscape is depicted prior to the application of the map in Eqs.~(\ref{eq:theta_map}) and (\ref{eq:phi_map}) indicating a non-trivial $\phi$ dependence. In (b) the map has been applied resulting in a manifestly $\widetilde{\phi}$ independent energy landscape.}
\end{figure}

The presence of a finite spin-orbit coupling term breaks the spin $SO(3)$ symmetry and introduces a preferred direction in the model. We study a Rashba-type spin-orbit coupling due to its relevance for systems with adatoms deposited on surfaces of bulk systems. The aim is to understand the circumstances under which the spin-orbit coupling can be gauged away and the effect absorbed into the spiral magnetic order. We note that for substrates with dimensionality greater than one such a transformation cannot be achieved exactly due to the presence of multiple non-commuting Pauli matrices in the Hamiltonian Eq.~\ref{eq:ham_app}. The additional SOC-term we consider is
\begin{eqnarray}
	\mathcal{H}_{\text{SO}} &=& t_{\text{so}} \sum_{\substack{i \\ \alpha\beta}}i c^{\dagger}_{i\alpha}\sigma^{x}_{\alpha\beta}c_{i+\delta_{y}\beta} \nonumber \\ && \qquad \qquad - ic^{\dagger}_{i\alpha}\sigma^{y}_{\alpha\beta}c_{i+\delta_{x}\beta} + \text{h.c.}\,, \label{eq:ham_app}
\end{eqnarray}
and once again consider two adatoms placed on the substrate a certain distance apart. The spin of one is kept fixed perpendicular to the plane and we use the Ansatz
\begin{eqnarray}
	\mbf{S}_2 = S\begin{pmatrix}
		\sin \theta \cos \phi \\
		\sin \theta \sin \phi \\
		\cos \theta
	\end{pmatrix}
\end{eqnarray}
to describe the other. Here $\phi$ describes the azimuthal, and $\theta$ the polar angle with respect to the first spin.
The total energy is evaluated for values of $\phi$ and $\theta$ corresponding to $165$ distinct points on a sphere, and the corresponding energy landscape is mapped out in Fig.~\ref{fig:tso_landscape}(a). At a glance, the energy landscape indicates a non-trivial dependence on the azimuthal angle, $\phi$. To understand if this is caused by the choice of rotation plane (and therefore can be gauged away), we consider the simple Hamiltonian
\begin{eqnarray}
	\mathcal{H} = &-&2(\cos k_x a + \cos k_y a) \nonumber \\ &+& \alpha (\sigma_y \sin k_x a - \sigma_x \sin k_y a) -\mu\,,
\end{eqnarray}
and observe that, when the spin-orbit coupling is weak, the modification to the Green function can be approximated as $G(x)\approx G_0(x) e^{-\frac{i}{2} t_{\text{so}}x \sigma_y}$, where $G_0(x)$ is the electron Green function at vanishing SOC~\cite{lutchyn10,heimes1,imamura04}. An evaluation of the RKKY exchange interaction reveals that it still contains only a term proportional to the angle between the two spins, $\mbf{S}_1 \cdot \widetilde{\mbf{S}}_2=\cos \widetilde{\theta}$, where the tilde refers to a new frame, related to the old frame \textit{via}
\begin{eqnarray}
	\widetilde{\theta} &=& \arccos \left( \cos \phi \sin \theta \sin \alpha a_{\rm ad} + \cos \theta \cos \alpha a_{\rm ad} \right)\,,\label{eq:theta_map} \\
	\widetilde{\phi} &=& \arctan \left( \frac{\sin \theta \sin \phi}{\cos \phi \sin \theta \cos \alpha a_{\rm ad} - \cos \theta \sin \alpha a_{\rm ad} } \right)\,. \label{eq:phi_map}
\end{eqnarray}
Here $\alpha$ can be related to the lattice parameter $t_{\text{so}}$ through $\alpha = C t_{\text{so}}$ where $C$ is a constant. For weak spin-orbit coupling, $C\approx 1$. As the spin-orbit coupling $t_{\text{so}}$ is increased, $C$ is renormalized through higher order contributions to the relation between $G(x)$ and $G_0$, until the point where the approximation breaks down, and the effect of spin-orbit coupling can no longer be gauged away. In Fig.~\ref{fig:tso_landscape}(b) we show the energy landscape in the transformed frame for $t_{\text{so}}=0.1$ and $J=0.1$, in which it is clear that the energy does not depend on $\widetilde{\phi}$.
Thus, as long as spin-orbit coupling is weak, its effect can be included as an additional pitch of the order along the magnetic chain.


\end{document}